\documentclass[a4paper,11pt]{article}

\usepackage[utf8]{inputenc}
\usepackage{amsmath}
\usepackage{amsfonts}
\usepackage{amssymb}
\usepackage{graphicx}
\usepackage[labelformat=empty]{subfig}
\usepackage[left=3cm,right=3cm,top=2.5cm,bottom=2.5cm]{geometry}

\usepackage{pstricks}
\definecolor{gray1}{gray}{0.1}
\definecolor{gray2}{gray}{0.2}
\definecolor{gray3}{gray}{0.3}
\definecolor{gray4}{gray}{0.4}
\definecolor{gray5}{gray}{0.5}
\definecolor{gray6}{gray}{0.6}
\definecolor{gray7}{gray}{0.7}
\definecolor{gray8}{gray}{0.8}
\definecolor{gray9}{gray}{0.9}

\begin{document}

\newcommand{\br}{\bar r}
\newcommand{\bQ}{\bar Q}
\newcommand{\bL}{\bar L}
\newcommand{\bq}{\bar q}

\title{Motion of test particles in a regular black hole space--time}
\author{Alberto Garc\'{\i}a$^1$, Eva Hackmann$^2$, Jutta Kunz$^3$, \\ Claus L\"ammerzahl$^{2,3}$ and Alfredo Mac\'{\i}as$^4$ \\
$^1$ Departamento de F\'{\i}sica, CINVESTAV-IPN, A.P. 14-740, M\'exico D.F., M\'exico \\
$^2$ ZARM, University of Bremen, Am Fallturm, 28359 Bremen, Germany \\
$^3$ Institute for Physics, University Oldenburg, 26111 Oldenburg, Germany \\
$^4$ Departamento de F\'{\i}sica, Universidad Autonoma Metropolitana--Iztapalapa, \\
A.P. 55-534, M\'exico D.F. 09340, M\'exico}

\maketitle

\begin{abstract}
We consider the motion of test particles in the regular black hole space-time given by Ay\'{o}n-Beato and Garc\'\i{}a in Phys.~Rev.~Lett.~80:5056 (1998). The complete set of orbits for neutral and weakly charged test particles is discussed, including for neutral particles the extreme and over-extreme metric. We also derive the analytical solutions for the equation of motion of neutral test particles in a parametric form and consider a post-Schwarzschild expansion of the periastron shift to second order in the charge.
\end{abstract}

\section{Introduction}

In the context of General Relativity (GR) many observable effects can be predicted which are not present in Newtonian physics. Einstein showed that the 'anomalous' shift of Mercury's perihelion is a relativistic effect, and also the bending of light of distant stars in the gravitational field of the Sun was predicted correctly by GR.
Most of the exact black hole solutions of the Einstein field equations possess a curvature singularity, e.g.~the Schwarzschild and the Kerr solution. The physical nature of such singularities is not really understood and, by the cosmic censorship conjecture, they should be hidden behind an event horizon so that the physics outside is not influenced by the singularity.

Assuming that the gravitational field of a gravitating object is described by the Reissner-Nordstr\"om solution, then its charge has an additional effect on the periapsis shift of neutral and charged particles \cite{Jaffe1922,Gackstatter1983,TeliPalaskar1984,Chaliasos2001,Hackmannetal08hd,GrunauKagramanova11}. It is to be expected that the charge of regular black hole solutions will also couple to the motion of test particles. This influence will in general be different from the Reissner-Nordstr\"om case, in particular close to the gravitating object where the two metrics strongly deviate. Therefore, the observation of the motion of massive test particles and of light near the central object is a very useful tool for extracting information about the nature of the gravitational field.

In 1968 Bardeen was the first to propose a regular model which avoids such problems \cite{Bardeen1968, Borde1994, Borde1997}. Others picked up that idea and a number of models was created which are commonly known as 'Bardeen black holes' \cite{BarrabesFrolov1996,Marsetal1996,CaboAyon-Beato1999}. However, these models were not obtained as exact solutions of the Einstein field equation coupled to some known physical sources. Other black hole models with regular center exist \cite{bron2,Dymni92}. Nevertheless, such models, although they satisfy the condition $T^0{}_0 = T^1{}_1$ of spherical symmetry, cannot be derived from a Lagrangian in general relativity \cite{Dymni92}. Another kind of them involves exotic scalar matter, i.e., scalar matter whose kinetic term is negative \cite{bron2}.

In 1998 Ay\'{o}n-Beato and Garc\'\i{}a found an exact solution to the Einstein equation coupled to a nonlinear electrodynamics with a physically reasonable source. This solution does not possess any curvature singularity and, therefore, is regular in this sense \cite{AyonBeatoGarcia98,AyonBeato:1999ec,AyonBeato:1999rg,Burinskii02}. In the following, we call this the Ay\'{o}n-Beato--Garc\'\i{}a space-time. It is uniquely described by two parameters: the mass $M$ and the electric charge $Q$. Asymptotically it approaches the Reissner-Nordstr\"om space-time, and for $Q=0$ it reduces to the Schwarzschild space-time. Later on it was shown by Ay\'{o}n-Beato and Garcia that the Bardeen model is indeed a solution of the Einstein equation coupled to a nonlinear magnetic monopole \cite{Ayon-BeatoGarcia2000}. Recently, various rotating regular black hole models have been presented \cite{Toshmatovetal14,ZilongBambi13,BambiModesto13,LiBambi14}. However, it is still under debate whether one of them can be interpreted as a solution of the Einstein field equations coupled to a nonlinear electrodynamics, see e.g. \cite{Azreg14}.

In this paper, we will analyze the motion of massive test particles in the Ay\'{o}n-Beato--Garc\'\i{}a space-time. We will use analytical methods to completely characterize the motion of uncharged and weakly charged massive test-particles. In addition, we will present the analytical solution for the equation of motion of neutral test particles.

For most of the common space-times, e.g.~Schwarzschild, Reissner-Nordstr\"om, or Kerr, analytical solutions to the equations of motions can be given in terms of elliptic functions. The motion of test-particles in Schwarzschild space-time was extensively discussed by Hagihara in 1931 \cite{Hagihara31} using Weierstrass elliptic functions. Analytical solutions for bound timelike geodesics in Reissner-Nordstr\"om space-time were given by Gackstatter \cite{Gackstatter1983} and for general timelike geodesics by Slez\'{a}kov\'{a} \cite{Slezakova06} in terms of Jacobian elliptic functions and integrals. In \cite{Hackmann10} analytical solutions for general timelike geodesics were given in terms of Weierstrass elliptic functions. A complete account on the motion of (magnetically and electrically) charged particles was given recently by Grunau and Kagramanova \cite{GrunauKagramanova11}, see also \cite{Hackmannetal08hd}. References to further work on geodesics in Schwarzschild and Reissner-Nordstr\"om space-time may be found in \cite{Sharp79}. A compact treatment of motion in Schwarzschild and Reissner-Nordstr\"om space-time is given in the book of Chandrasekhar \cite{Chandrasekhar83}. For the regular Bardeen model, the motion of massive test particles was discussed by Zhou et al \cite{Zhouetal2012} and gravitational lensing by Eiroa and Sendra \cite{EiroaSendra2011}. Note that an interesting aspect of solutions in non-linear electrodynamics is the fact that photons propagate along the null geodesics of an effective metric instead of the space-time metric \cite{Plebanski1970,Gutierrez:1981ed,Novello:1999pg}. For the Ay\'{o}n-Beato--Garc\'\i{}a space-time the propagation of photons has been studied by Novello et al.~\cite{Novello:2000km}. The effective metric for photons is quite complex and in particular exhibits singularities at finite distances from the center.

It is clear that the study of orbits around charged black holes is important from conceptual and theoretical points of view. However, since it is the general belief that astrophysical black holes possess a small or null charge the astrophysical importance of such studies are considered to be of limited applications. However, there are some discussions in the situation that black holes embedded in an external (e.g. galactic) magnetic field and surrounded by a plasma may accrete some charge \cite{Wald74,RuffiniWilson75,Damouretal78}. Despite of this limitation, there are quite a few studies of astrophysical implications of charged black holes \cite{MisraLevin10}. The orbits of neutral and charged point particles around charged Reissner-Nordstr\"om and Kerr-Newman black holes has been completely given by \cite{GrunauKagramanova11,Hackmannetal2013}, and the circular orbits in these space-times have been extensively discussed in \cite{Puglieseetal2013,Puglieseetal2011}. In addition, the orbits in extreme Reissner-Nordstr\"om dihole space-times have been presented in \cite{Wuenschetal13}. In \cite{MisraLevin10} rational orbits and their influence on the creation of gravitational waves have been discussed (see also \cite{Zilhaoetal12,Zilhaoetal14}).

While the charge of black holes may modify the physics of matter in the vicinity of black holes, charged black holes within some nonlinear electrodynamics theory may also change essentially the properties of space and time \cite{AyonBeatoGarcia98, Ruffinietal2013}. Within such a framework, gravitational lensing by Einstein-Born-Infeld black holes has been calculated \cite{Eiroa06}, see also \cite{EiroaSendra2011} which is applied to observations at the supermassive black hole at the center of our galaxy. A further motivation for our study could be that, since there is no singularity at $r = 0$, a material source might exist for this geometry that could evolve from one asymptotically flat region to another. It is known that such sources do not exist for the Reissner-Nordstr\"om black hole because shell crossings block the passage through the throat between the $r = 0$ singularities, see \cite{Ori90,KrasinskiBolejko06,KrasinskiBolejko07}.

It is also clear that the charge of black holes either for singular or regular black holes will be of importance in the understanding of physics of the accretion of plasma and the creation of jets.

The outline of the paper is as follows: First we review some general properties of the Ay\'{o}n-Beato--Garc\'\i{}a space--time and derive the equations of motions, in the second and third sections. In the fourth section we discuss all types of orbital motion for neutral test particles, including the over-extreme case without horizons. We also analyze the stability of circular orbits and show the position of the innermost stable circular orbit as a function of the charge $Q$. The special case of weakly charged particles in a black hole space-time is also considered. In the fifth section we derive the analytical solution to the equation of motion for neutral test particles in terms of Weierstrass elliptic functions. For this we introduce a new affine parameter analogously to the Mino time \cite{Mino03}. Finally, we discuss the periastron shift for neutral test particles and close with a summary

\section{The Ay\'{o}n-Beato--Garc\'\i{}a space-time}

In this section we review some of the basic properties of the Ay\'{o}n-Beato--Garc\'\i{}a space-time, which is regular in the sense that it has no curvature singularity. This regularity is achieved by coupling to a general model of non-linear electrodynamics.

\subsection{The action}

To obtain electrically charged solutions of the Einstein-Pleba\'nski class of non-linear electrodynamics \cite{Plebanski1970,Burinskii02,GHLM2012} equations one starts from the action
\begin{equation}
S = \int d^4 x \left[ \frac{1}{16 \pi} R - \frac{1}{4 \pi} {\cal{L}}(F) \right]
\, ,
\label{eq:action}
\end{equation}
where $R$ is the scalar curvature and $\cal{L}$ is a function of
$F= \frac{1}{4} F_{\mu\nu} F^{\mu\nu}$
\cite{AyonBeatoGarcia98}.
One can also describe the system under consideration by means of
the function ${\cal H}(P)$ obtained from the following Legendre
transformation \cite{Plebanski1970,Salazar:1987ap}
\begin{equation}
{\cal H}\equiv 2F{\cal L}_F-{\cal L}.  \label{eq:Leg}
\end{equation}
Defining
\begin{equation}
P_{\mu \nu }\equiv {\cal L}_FF_{\mu \nu },
\end{equation}
it follows that ${\cal H}$ is a function of
\begin{equation}
P\equiv \frac 14P_{\mu \nu }P^{\mu \nu }=({\cal L }_F)^2F.
\end{equation}

The specific function ${\cal H}$ employed for the regular
black hole solution \cite{AyonBeatoGarcia98}
is given as
\begin{equation}
{\cal H}(P)=P\,\frac{\left( 1-3\sqrt{-2\,Q^2P}\right) }{\left( 1+\sqrt{%
-2\,Q^2P}\right) ^3}-\frac 3{2\,Q^2s}\left( \frac{\sqrt{-2\,Q^2P}}{1+\sqrt{%
-2\,Q^2P}}\right) ^{5/2},  \label{eq:H}
\end{equation}
where $s=|Q|/2m$ and the invariant $P$ is a negative quantity,
\begin{equation}
P = - \frac{Q^2}{2 r^4} ,
\end{equation}
where the integration constant $Q$ plays the role of the electric charge.
The components of $P_{\mu\nu} = {\cal L}_F F_{\mu\nu}$ are just the
electromagnetic field excitations ${\bf D}$ \cite{Burinskii02,GHLM2012}.

\subsection{The metric and the electric field}

The metric of the regular spherically symmetric space-time we are considering is given by \cite{AyonBeatoGarcia98}
\begin{equation}
ds^2 = g_{tt} dt^2 - g_{rr} dr^2 - r^2 \left(d\vartheta^2 + \sin^2\vartheta d\varphi^2\right)\,,  \label{generalmetric}
\end{equation}
where
\begin{equation}
g_{tt} = \frac{1}{g_{rr}} = 1 - \frac{2 M r^2}{\left(r^2 + Q^2\right)^{\frac{3}{2}}} + \frac{Q^2 r^2}{\left(r^2 + Q^2\right)^2} \, .
\end{equation}
Asymptotically, that is for $r \rightarrow \infty$, this metric behaves as a Reissner--Nordstr\"om metric.
At the center, $r=0$, the metric is regular
\begin{equation}
g_{rr} \approx 1 - \frac{1}{3} \Lambda r^2\,, \quad \Lambda=3\frac{|Q|-2}{|Q|^3}\,.
\end{equation}

The electric field is given by
\begin{equation}
E_r=Q\,r^4 \left( \frac{r^2-5\,Q^2}{(r^2+Q^2)^4}+\frac{15}{2}\,
\frac{M}{(r^2+Q^2)^{\frac72}} \right)  \label{eq:E}
\end{equation}
and vanishes at the origin. As pointed out in \cite{Novello:2000km}, $E_r$ does not have finite positive zeros as long as $|Q|<\frac{3}{2}M$. Above that value $E_r$ possesses one finite zero. Also, for $Q>2M$ the energy density becomes negative for small values of $r$.

\subsection{The horizons}

\begin{figure}
\includegraphics[width=0.45\textwidth]{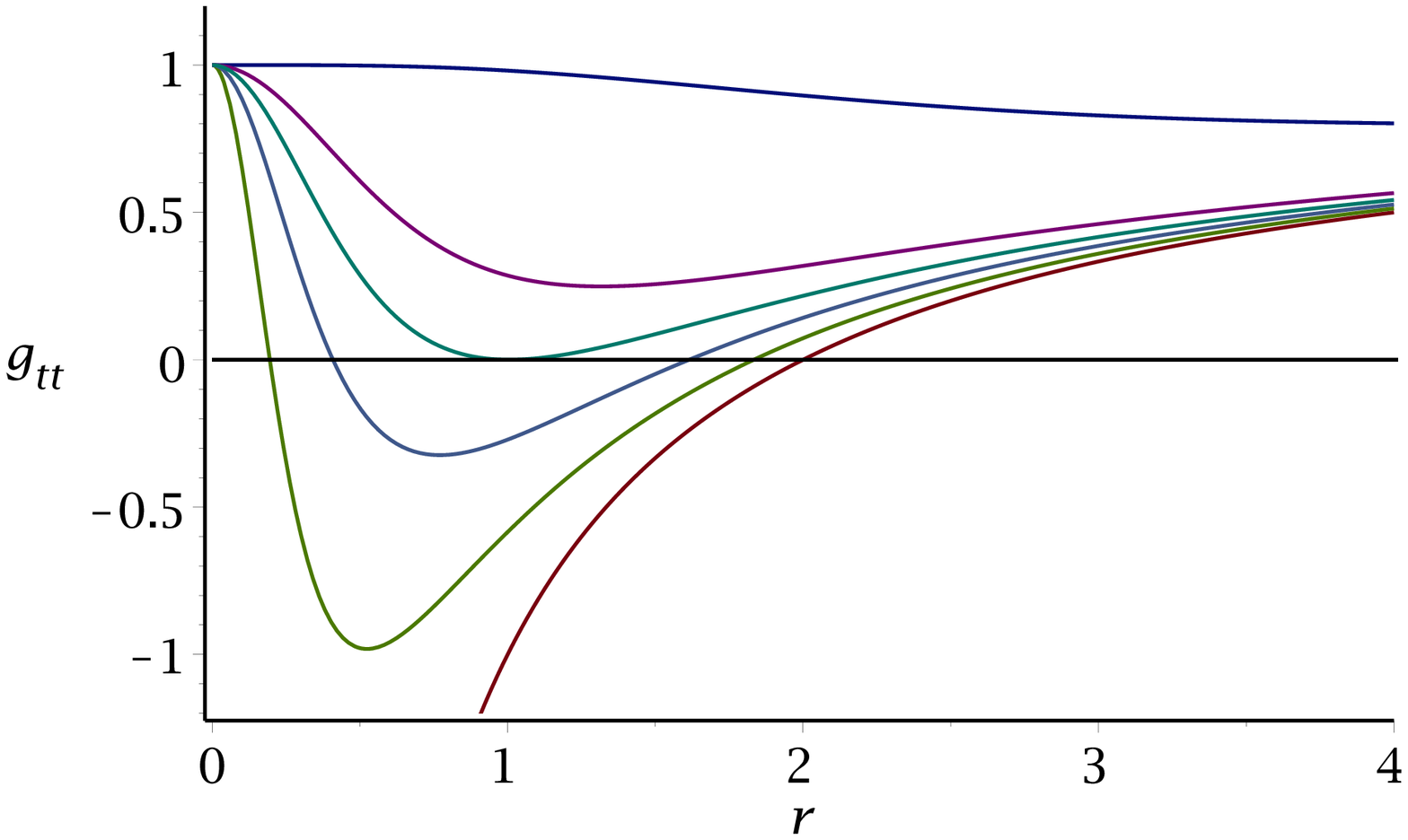}\quad
\includegraphics[width=0.45\textwidth]{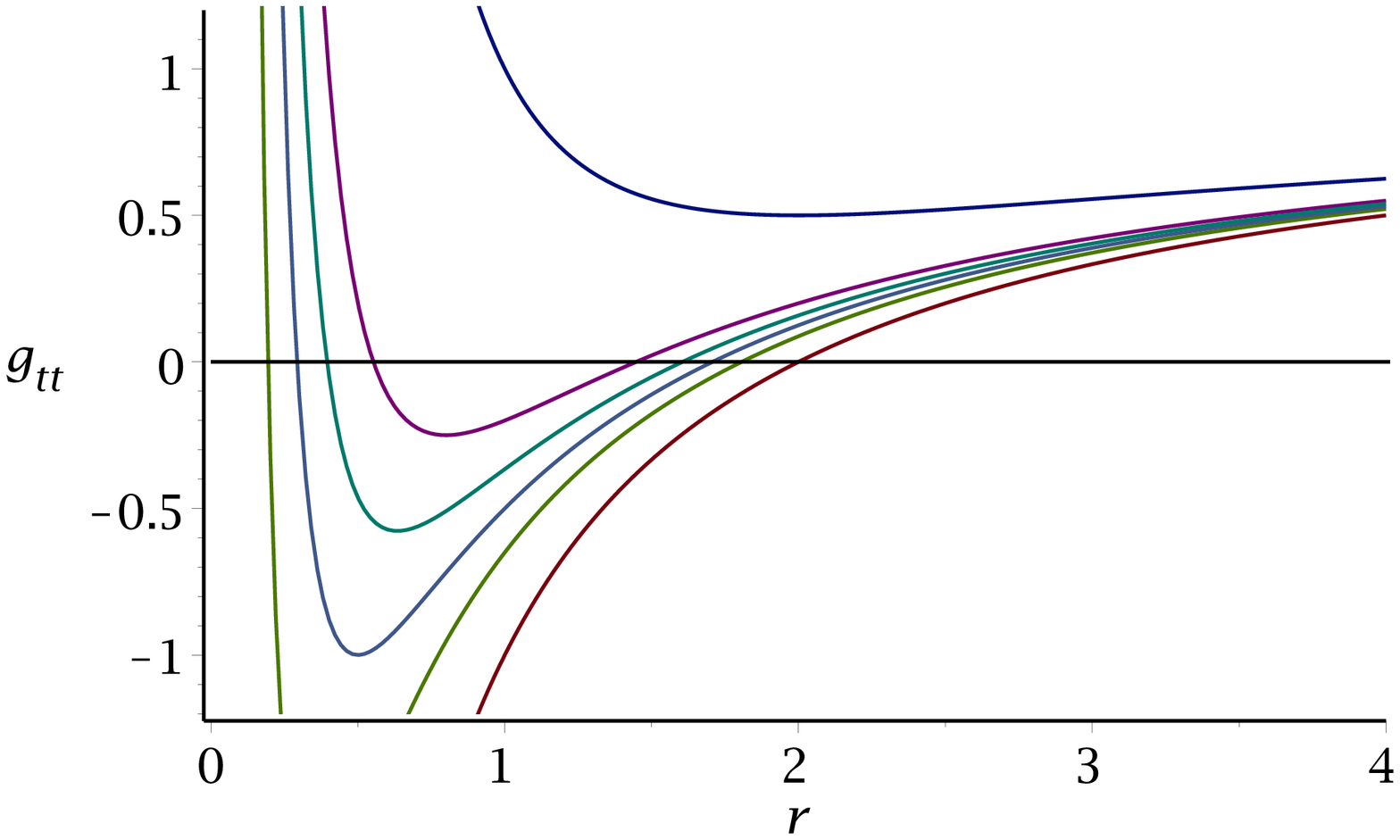}
\caption{Left: The metric function $g_{tt}$ of the Ay\'{o}n-Beato--Garc\'\i{}a space--time for the parameters $Q = 0$, $0.035M$, $0.05M$, $Q_{\rm crit}$, $0.8M$, and $2M$ (from bottom to top). 
Right: For comparison, the metric function $g_{tt}$ for Reissner--Nordstr\"om space--times. For large $r$ both metrics approach each other, for small $r$ the Reissner--Nordstr\"om metric tends to infinity yielding a singularity while the Ay\'{o}n-Beato--Garc\'\i{}a metric approaches $g_{tt}=1$.}
\label{Fig:gtt}
\end{figure}

The horizons are given by the vanishing of the metrical coefficient $g_{tt} = 0$, i.e.~by the real solutions of
\begin{align}
r^8+2(3Q^2-2M^2)r^6+Q^2(11Q^2-4M^2)r^4+6Q^6r^2+Q^8=0\,.
\end{align}
As $r$ and $Q$ appear only quadratically it suffices for solving this equation to consider the positive solutions and $Q>0$. According to Descartes' rule of signs this polynomial has two or no positive zeros. The above expression has a double zero at $Q_{\rm crit} \approx 0.634 M$ which corresponds to the extremal case with a horizon at $r_{\rm crit}\approx 1.005 M$. For $Q<Q_{\rm crit}$ there are two horizons, for $Q>Q_{\rm crit}$ there are no horizons. If $Q=0$ the Schwarzschild case $r=2M$ is recovered. As in Reissner--Nordstr\"om space-times the electromagnetic field acts repulsive in the metrical sector. Examples of $g_{tt}$ are shown in Fig.~\ref{Fig:gtt}.

\begin{figure}[t]
\includegraphics[width=\textwidth]{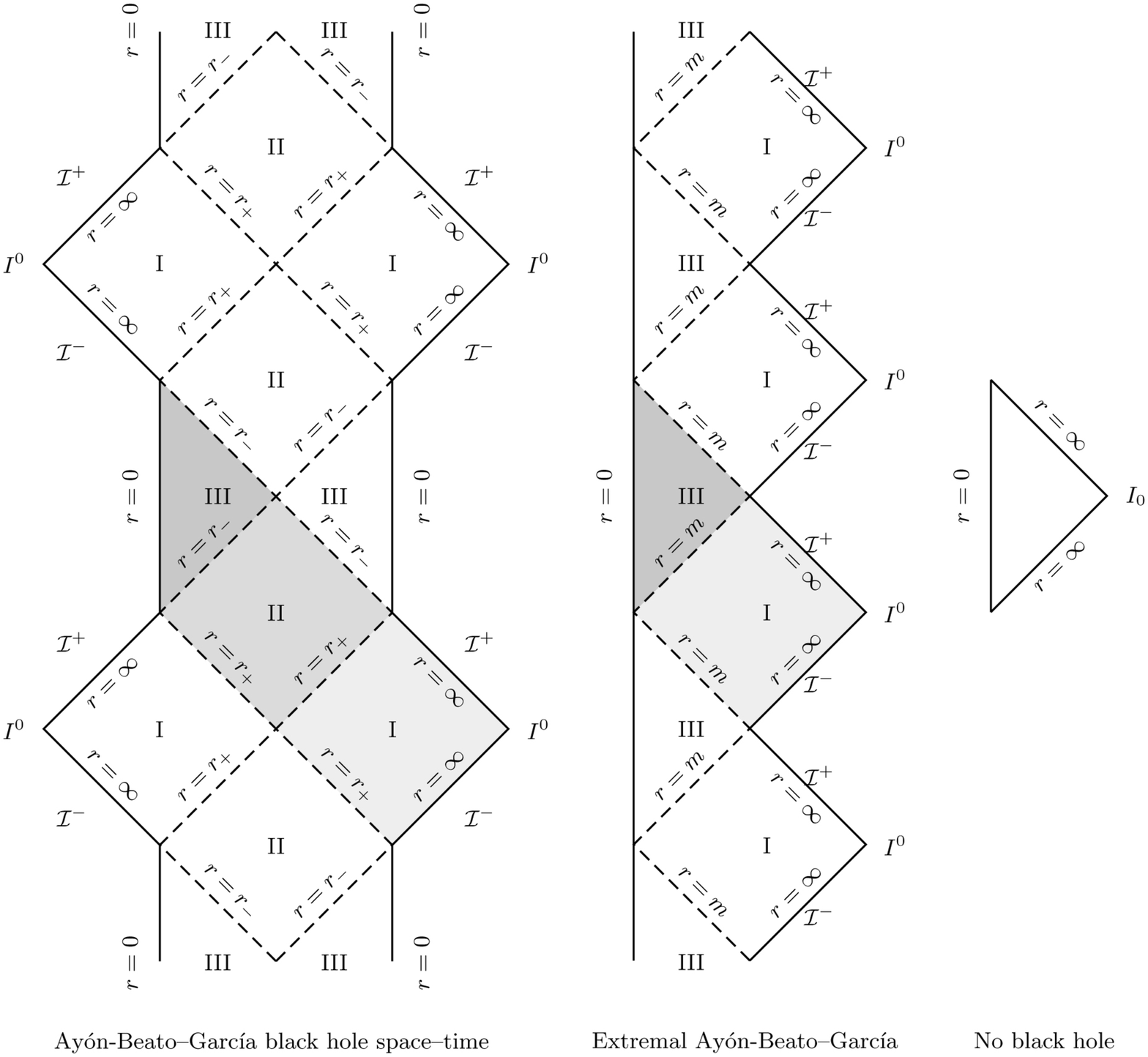}
\caption{The Carter--Penrose diagrams of the various types of Ay\'{o}n-Beato--Garc\'\i{}a space--times: (a) black hole, (b) extremal black hole, (c) no black hole. Please note that the vertical lines $r=0$ do not represent a singularity as in Reissner-Nordstr\"om space-times but, instead, represent a regular part of the space-time analogous to the origin in Minkowski space-time. \label{CPABG}}
\end{figure}

In the case of a black hole space--time we have two horizons. The corresponding Carter--Penrose diagrams of the black hole space-time and the extreme and over-extreme cases are shown in Fig.~\ref{CPABG}. The diagrams look very similar to the Reissner-Nordstr\"om case except that in our case the vertical lines $r=0$ do not indicate a singularity but represent simply a regular part of the space-time analogous to the spatial origin $r=0$ in Minkowski space-time. In particular, in the right diagram there are no horizons and, thus, there is no black hole, and the space-time possesses the topology of Minkowski space-time.

\section{The equations of motion}

The equations of motion for (charged) test particles are given by
\begin{equation}
q F^\mu{}_\nu \frac{dx^\nu}{ds} = \frac{d^2x^\mu}{ds^2} + \left\{\begin{smallmatrix} \mu \\ \rho\sigma\end{smallmatrix}\right\} \frac{dx^\rho}{ds} \frac{dx^\sigma}{ds}
\end{equation}
where $F$ is the electromagnetic field strength ($F_{tr}=E_r$, $F_{\mu\nu}=0$ else), $s$ the proper time, $q$ is the specific charge, and $\left\{\begin{smallmatrix} \mu \\ \rho\sigma\end{smallmatrix}\right\} = \frac{1}{2} g^{\mu\nu} \left(\partial_\rho g_{\sigma\nu} + \partial_\sigma g_{\rho\nu} - \partial_\nu g_{\rho\sigma}\right)$. Since the space-time and the electromagnetic field are spherically symmetric we may restrict without loss of generality the motion to the equatorial plane $\theta=\pi/2$.

\subsection{Neutral test particles}
Let us first consider the case $q=0$. For the metric above we have two conserved quantities, energy and angular momentum
\begin{equation}
E = g_{tt} \frac{dt}{ds} \, , \qquad L = r^2 \frac{d\varphi}{ds} \, .
\end{equation}
In addition, $g_{\mu\nu} \frac{dx^\mu}{ds} \frac{dx^\nu}{ds} = \epsilon$, where $\epsilon = 1$ for massive point particles and $\epsilon = 0$ for null geodesics (which correspond to the high energy limit of massive test particles and are different from photon orbits). Then the geodesic equation gives the following ordinary differential equations
\begin{align}
\left(\frac{dr}{ds}\right)^2 & = \frac{1}{g_{tt} g_{rr}} \left(E^2 - g_{tt} \left(\epsilon + \frac{L^2}{r^2}\right)\right)\,, \\
\left(\frac{dr}{d\varphi}\right)^2 & = \frac{r^4}{L^2} \frac{1}{g_{tt} g_{rr}} \left(E^2 - g_{tt} \left(\epsilon + \frac{L^2}{r^2}\right)\right)\,, \label{drdphi}\\
\left(\frac{dr}{dt}\right)^2 & = \frac{1}{E^2} \frac{g_{tt}}{g_{rr}} \left(E^2 - g_{tt} \left(\epsilon + \frac{L^2}{r^2}\right)\right)\,.
\end{align}
In our case $g_{tt} g_{rr} = 1$ and with the dimensionless quantities $\bar s=s/M$, $\bar r=r/M$, $\bar Q=Q/M$, and $\bL=L/M$ the first equation reduces to
\begin{align}
\left(\frac{d\br}{d\bar s}\right)^2 & =  E^2-\left( 1-\frac{2\br^2}{(\br^2+\bQ^2)^{\frac{3}{2}}}+\frac{\bQ^2\br^2}{(\br^2+\bQ^2)^2} \right)\left(\epsilon+\frac{\bL^2}{\br^2}\right) =:R(\br)\,, \label{drds}
\end{align}

From \eqref{drds} we read off an effective potential
\begin{equation}
V_{\rm eff} = g_{tt} \left(\epsilon + \frac{L^2}{r^2}\right) -\epsilon = \left( 1-\frac{2\br^2}{(\br^2+\bQ^2)^{\frac{3}{2}}}+\frac{\bQ^2\br^2}{(\br^2+\bQ^2)^2} \right)\left(\epsilon+\frac{\bL^2}{\br^2}\right) -\epsilon\, .
\end{equation}
From the square on the left hand side of \eqref{drds} it is necessary that $E^2-\epsilon\geq V_{\rm eff}$ for a solution to exist. This is equivalent to $R(\br)\geq0$. Instead of discussing the effective potential we will consider this condition later on. For $\bQ = 0$ we recover the usual Schwarzschild effective potential.

In order to get rid of the square root appearing in the equation \eqref{drds} we introduce a new variable $u=1/\sqrt{\br^2+\bQ^2}$. This restricts $u$ to $0\leq u \leq \bQ^{-1}$ and simplifies the equation to
\begin{align}
\left(\frac{du}{d\bar s}\right)^2 & = u^4 P_6(u) \label{duds}\,,
\end{align}
where $P_6$ is a polynomial of degree six,
\begin{align}
P_6(u)= \bQ^2\alpha u^6 -2\alpha u^5 -\bQ^2\beta u^4 +2\beta u^3 -(E^2\bQ^2+\bL^2)u^2 +2u\epsilon +E^2-\epsilon\,,
\end{align}
with $\alpha=\bQ^2(\bL^2-\epsilon\bQ^2)$, $\beta=(\bL^2-2\epsilon\bQ^2)$. Written as an integral equation this reads
\begin{align}
\bar s-\bar{s}_0 = \int_{u_0}^u \frac{du}{u^2\sqrt{P_6(u)}}\,, 
\end{align}
where $u(\bar{s}_0)=u_0$
are the initial values. The right hand side is a hyperelliptic integral of the third kind. We will analytically solve this equation in a parametric form later on.

\subsection{Charged test particles}
Now consider the case $q\neq0$. The electrostatic potential is
\begin{equation}
A_0 = - Q \frac{r^5}{(r^2 + Q^2)^3} + \frac{3}{2} \frac{M}{Q} \frac{r^5}{(r^2 + Q^2)^{\frac52}}\,.
\end{equation}
Here we can not perform the limit $Q \rightarrow 0$ which is no problem since the potentials have no physical meaning. For the corresponding field strength this problem does not occur. However, the fact that $A_0$ is an odd function of $r$ has strong consequences for the effective equations of motion, as we will below.

The constants of motions are modified due to the charge to
\begin{align}
E & = u_\mu \xi^\mu_{(t)} + q A_0 = g_{tt} \frac{dt}{ds}\,, \\
L & = - u_\mu \xi^\mu_{(\varphi)} = - g_{\varphi\varphi} \frac{d\varphi}{ds} = r^2 \frac{d\varphi}{ds}\,, \\
1 & = g_{\mu\nu} u^\mu u^\nu = g_{tt} \left(\frac{dt}{ds}\right)^2 - g_{rr} \left(\frac{dr}{ds}\right)^2 - g_{\varphi\varphi} \left(\frac{d\varphi}{ds}\right)^2\,,
\end{align}
where $u$ is the four-velocity and $\xi_{(t)}$ and $\xi_{(\varphi)}$ are the Killing vectors, resulting in
\begin{equation}
\left(\frac{dr}{ds}\right)^2 = \left(E - q A_0\right)^2 - g_{tt} \left(1 + \frac{L^2}{r^2}\right)\,. \label{EOMCharge1}
\end{equation}
If we use the dimensionless quantities introduced above and $\bq=q\bQ^{-1}$ this can be written as
\begin{align}
\left(\frac{d\br}{d{\bar s}}\right)^2 & = R(\br) + \bq E \left(\frac{2\bQ^2\br^5}{(\br^2 + \bQ^2)^3} - \frac{3\br^5}{(\br^2 + \bQ^2)^{\frac52}}\right)\nonumber\\
& \quad + \bq^2 \left(\frac{\bQ^4\br^{10}}{(\br^2 + \bQ^2)^6} - \frac{3\bQ^2\br^{10}}{(\br^2 + \bQ^2)^{\frac{11}{2}}} + \frac{9}{4} \frac{\br^{10}}{(\br^2 + \bQ^2)^5} \right) =: R_{\bq}(\br)\,, \label{chargedrds}
\end{align}
with $R(\br)$ as in \eqref{drds} and $R_{\bq=0}=R$. With the substitution $u=1/\sqrt{\br^2+\bQ^2}$ as above this simplifies to
\begin{align}
\left(\frac{du}{d{\bar s}}\right)^2 & = u^4 \big[ P_6(u) + \bq E (2\bQ^2u-3)(1-\bQ^2u^2)^{\frac72} \nonumber\\
& \qquad + \bq^2 \left(\bQ^4u^2-3\bQ^2u+\frac{9}{4}\right) (1-\bQ^2u^2)^6 \big] =: u^4U(u) \label{chargeduds}
\end{align}
with $P_6$ as in \eqref{duds}. Unfortunately, this substitution does not eliminate all roots, which apparently cannot be avoided for charged particles. This type of equation has the same structure as the equation of motion for the Kehagias-Sfetsos black hole in Ho\v{r}ava-Lifshitz gravity \cite{EHKKLS11_num}. To our knowledge, an analytical solution is not known. However, the topology of orbits can be analyzed, which will be done below for weakly charged test particles in black hole space-times.

\section{Types of orbits}

The equation of motion \eqref{drds} is invariant under changes in sign of $E$, $\bL$, and $\bQ$ whereas \eqref{chargedrds} is invariant under changes of sign of $\bq E$, $\bL$, and $\bQ$. Therefore, we choose $\bL>0$ and $\bQ>0$, as well as $E>0$ for neutral test particles and $\bq>0$ for charged particles. 
Due to the square on the left hand sides of \eqref{drds} and \eqref{chargedrds} a necessary condition for the existence of a solution is that $R(\br)\geq0$ and $R_{\bq}(\br)\geq0$. In the following we will discuss these conditions separately.

\subsection{Massive neutral test particles}

Let us consider whether $R(\br)\geq0$ is fulfilled at $\br=0,\infty$. First, we observe that $R(\br)\to E^2-\epsilon$ for $\br\to\infty$ which implies that only for $E^2\geq\epsilon$ infinity may be reached. For $\br\to0$, it is $R(\br)=-\frac{\bL^2}{\br^2} + E^2-\epsilon+\frac{\bL^2}{\bQ^3}(\bQ-2) + \mathcal{O}(\br^2)$ implying that $\br=0$ may only be reached for $\bL=0$ and $E^2\geq\epsilon$. Furthermore, in black hole space-times $R$ is always positive between the horizons due to $g_{tt}<0$ there.

Once we know the behavior of $R(\br)$ at the origin and at infinity, all possible types of orbits (i.e.~regions of $R(r)\geq0$) can be inferred from the number of positive real zeros. Suppose that for a given set of parameters all types of orbits have been identified. If now the constants of motion are varied, the number of real zeros of $R$ changes at those sets of parameters where double zeros occur, which correspond to circular orbits. Therefore, the analysis of circular orbits is the key element for identifying all possible types of orbits. For this analysis, we switch to the coordinate $u=1/\sqrt{\br^2+\bQ^2}$, see \eqref{duds}. Solving $P_6(u)=0$, $\frac{dP_6}{du}(u)=0$ for $E^2$ and $\bL^2$ yields
\begin{equation}\label{EL}
\begin{aligned}
E^2 & = \frac{1}{A(u)} (\bQ^4u^4- 2\bQ^2u^3 -\bQ^2u^2 +2u -1)^2\,,\\ 
\bL^2 & = \frac{1}{uA(u)} (2\bQ^4u^3 -3\bQ^2u^2 -\bQ^2u +1)(1-\bQ^2u^2)^2\,, 
\end{aligned}
\end{equation}
where $A(u):=2\bQ^6u^6-3\bQ^4u^5-4\bQ^4u^4+6\bQ^2u^3+2\bQ^2u^2-3u+1$. Note that for $u\to0$ the expression for $\bL^2$ diverges but $E^2=1-u+\mathcal{O}(u^2)$. Near $u=\bQ^{-1}$ we get $\bL^2=4(\bQ-2)\bQ^3(u-\bQ^{-1})^2 + \mathcal{O}((u-\bQ^{-1})^3)$, $E^2\to1$ and, therefore, circular orbits near $\br=0$ may exist for $\bQ>2$. In the limit $\bQ=0$ equations \eqref{EL} reduce to the well known Schwarzschild expressions $E^2=\frac{(1-2u)^2}{1-3u}=\frac{(r-2)^2}{r(r-3)}$, $\bL^2=\frac{1}{u(1-3u)}=\frac{r^2}{r-3}$.

\begin{figure}
\centering
\includegraphics[width=0.4\textwidth]{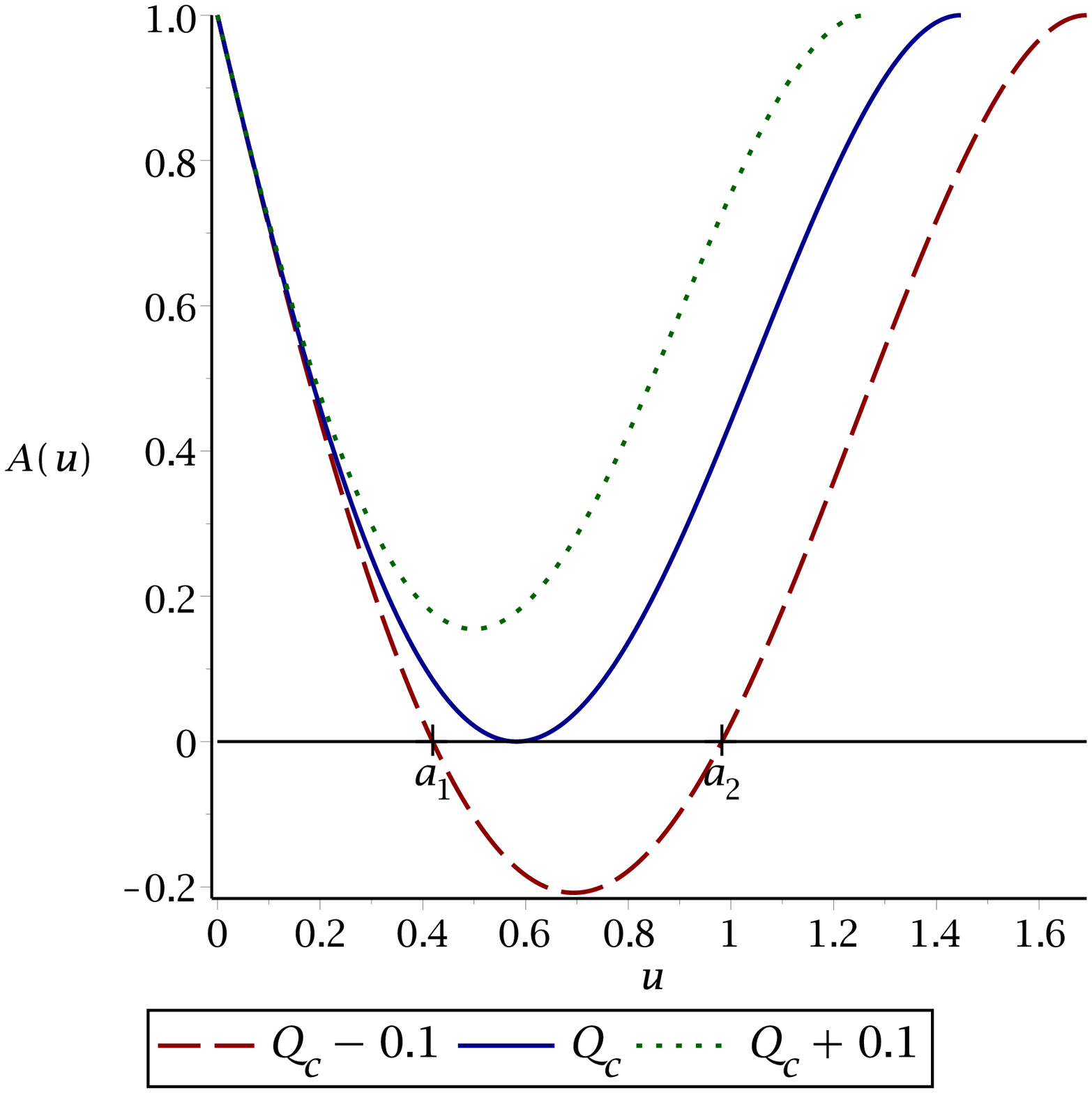}\qquad
\includegraphics[width=0.4\textwidth]{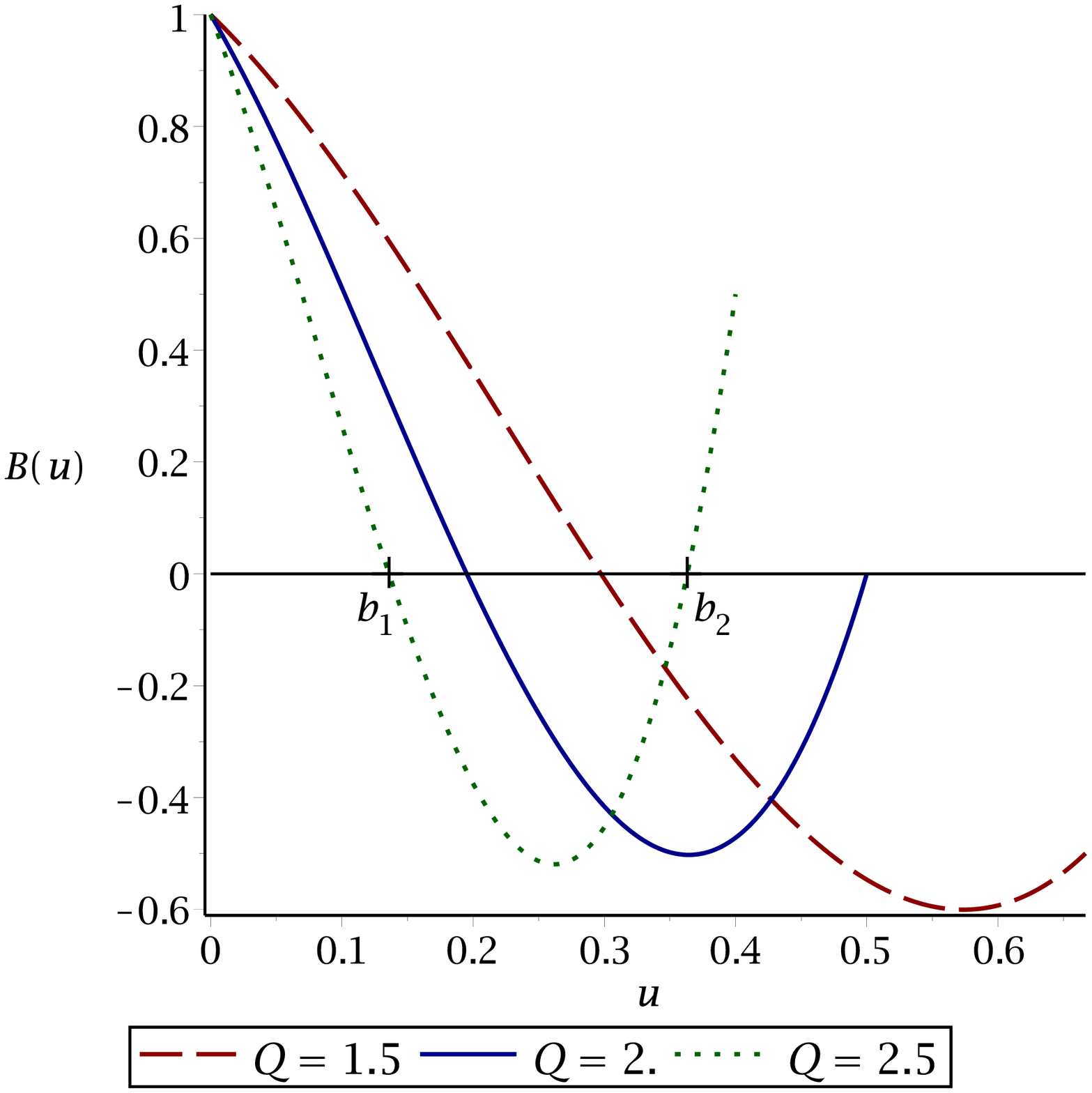}
\caption{Analysis of $A(u)$ and $B(u)$. Left: in the admissible range $A(u)$ has two zeros $a_1<a_2$ for $\bQ<\bQ_c$, a double zero at $\bQ=\bQ_c$ and $A(u)>0$ for all $u \in [0,\bQ^{-1}]$ if $\bQ>\bQ_c$. Right: in the admissible range, $B(u)$ has a single zero $b_1$ if $\bQ<2$ and two zeros $b_1<b_2$ if $\bQ\geq2$.}
\label{Fig:AB}
\end{figure}

From the equations \eqref{EL} two necessary conditions for the existence of circular orbits can be derived: the polynomial in the denominator of $E^2$ has to be positive, $A(u)>0$, and accordingly $B(u):=2\bQ^4u^3 -3\bQ^2u^2 -\bQ^2u +1\geq0$ is also necessary. Let us consider the first condition: $A(u)$ has a double zero in $[0,\bQ^{-1}]$ at approximately $\bQ_c \approx 0.691$, $u_c\approx 0.582$ ($\br_c\approx 1.572$). For $\bQ<\bQ_c$ there are two zeros $a_1<a_2$ in $[0,\bQ^{-1}]$ with positive $A$ for $0\leq u < a_1$ and $a_2 < u \leq \bQ^{-1}$. If $\bQ>\bQ_c$ then $A(u)$ is positive in the complete range $[0,\bQ^{-1}]$. Concerning the second condition, $B$ may have up to two positive real zeros (by Descartes' rule of signs) and $B(0)=1$, $B(\bQ^{-1})=\bQ-2$. From that we infer that for $\bQ<2$ the polynomial $B$ is positive in $0 \leq u \leq b_1 < \bQ^{-1}$ for the smallest positive zero $b_1$ of $B$. For a graphical summary of these findings see Fig.~\ref{Fig:AB}. As $A(b_1)<0$ if $\bQ<\bQ_{\rm crit}$, $b_1=a_2$ if $\bQ=\bQ_{\rm crit}$, and $A(b_1)>0$ with $b_1>a_2$ if $\bQ_{\rm crit}<\bQ<\bQ_c$ we conclude that $a_1<b_1<a_2$ if $\bQ<\bQ_{\rm crit}$ and $a_2\leq b_1$ if $\bQ_{\rm crit}\leq\bQ\leq\bQ_c$. Summarized,
\begin{itemize}
\item $0\leq\bQ<\bQ_{\rm crit}$: for all black hole space-times there may be circular orbits only for $0\leq u < a_1$ or, equivalently, $(a_1^{-2}-\bQ^2)^{\frac{1}{2}} < \br \leq \infty$;
\item $\bQ_{\rm crit}\leq\bQ\leq\bQ_c$: circular orbits exist in $0\leq u < a_1$ and $a_2< u \leq b_1$ or, equivalently, $(a_1^{-2}-\bQ^2)^{\frac{1}{2}} < \br \leq \infty$ and $(b_1^{-2}-\bQ^2)^{\frac{1}{2}} \leq \br < (a_2^{-2}-\bQ^2)^{\frac{1}{2}}$;
\item $\bQ_c\leq\bQ<2$: circular orbits exist for $0\leq u\leq b_1$ or, equivalently, $(b_1^{-2}-\bQ^2)^{\frac{1}{2}} \leq \br \leq \infty$;
\item $\bQ\geq2$: an additional inner region with circular orbits appears, $b_2\leq u < \bQ^{-1}$, where $b_2$ is the largest zero of $B$, i.e.~circular orbits exist in $(b_1^{-2}-\bQ^2)^{\frac{1}{2}} \leq \br \leq \infty$ and $0\leq \br\leq (b_2^{-2}-\bQ^2)^{\frac{1}{2}}$.
\end{itemize}

\begin{figure}
\centering
\includegraphics[width=0.4\textwidth]{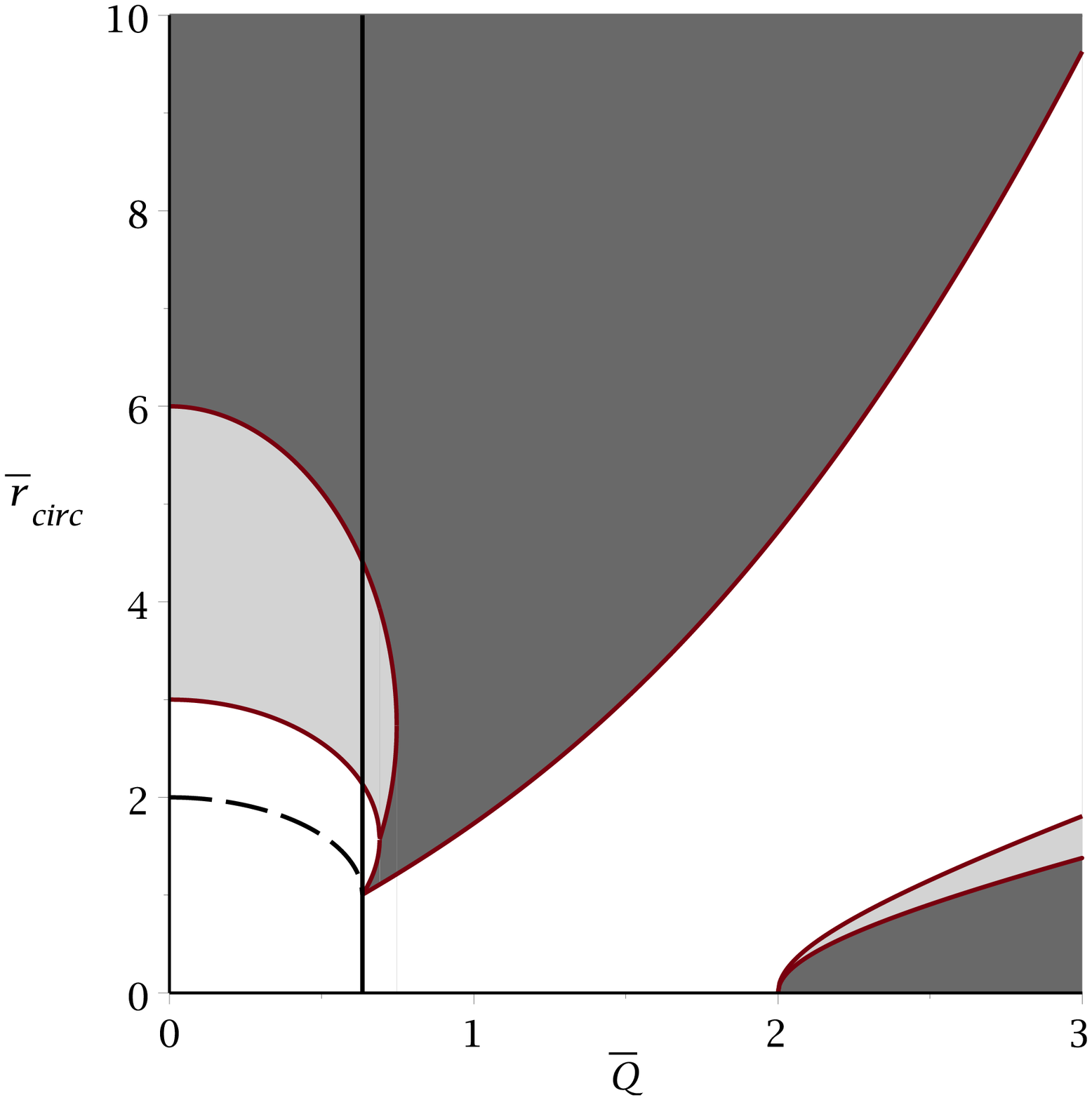}\qquad
\includegraphics[width=0.4\textwidth]{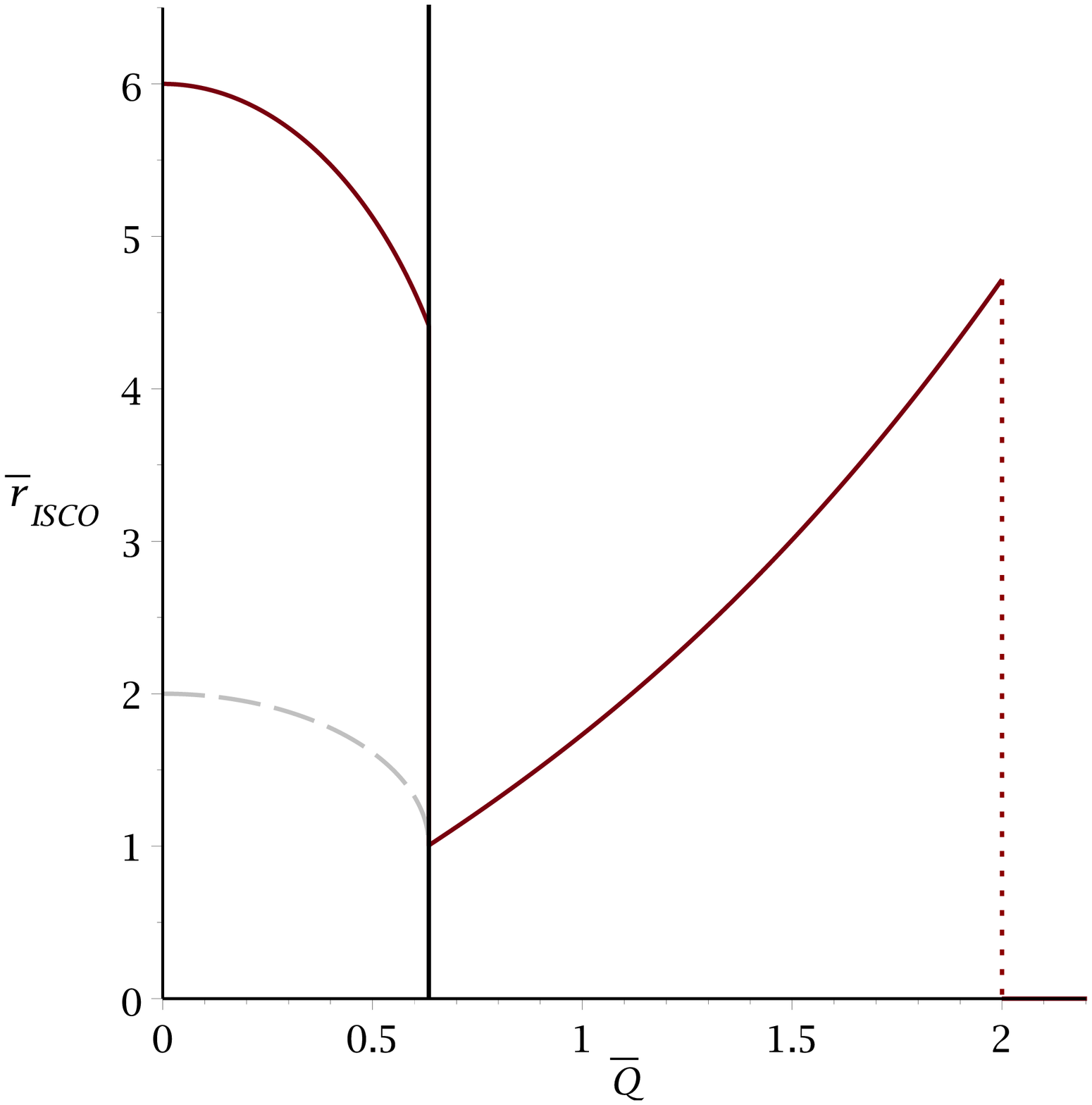}
\caption{Radii of circular orbits for neutral test particles in Ay\'{o}n-Beato--Garc\'\i{}a space-times as a function of the charge $\bQ$. The black vertical line marks $\bQ=\bQ_{\rm crit}$ and the grey dashed line the outer horizon. Left: The dark grey regions correspond to stable and the light grey to unstable circular orbits. In the white regions no circular orbits are possible. Right: Radius of the innermost stable circular orbit. The first jump occurs at $\bQ_{\rm crit}$ from $(a_1^{-2}-\bQ^2)^{\frac{1}{2}}$ to $(b_1^{-2}-\bQ^2)^{\frac{1}{2}}$ due to the vanishing horizons. At $\bQ=2$ an additional maximum in the effective potential appears which causes the second jump to $\br=0$.}
\label{Fig:ISCO}
\end{figure}

The stability of the circular orbits can be analyzed by considering the second derivative of $P_6$ (see eq.~\eqref{duds}). If $P_6$ has a maximum then the circular orbit is stable and else unstable. Inserting \eqref{EL} in $\frac{d^2P_6}{du^2}(u)$ yields
\begin{align}
\frac{d^2P_6}{du^2}(u) & = \frac{-2u^3}{A(u)} T(u)\,,\\
T(u) & = 4\bQ^{10}u^9 -9\bQ^8u^8 +6\bQ^6(1-2\bQ^2)u^7 +27\bQ^6u^6 -6\bQ^4(3-4\bQ^2)u^5 -42\bQ^4u^4 \nonumber\\
& \quad +6\bQ^2(3-2\bQ^2)u^3 +15\bQ^2u^2 -6u+1\,.
\end{align}
As $\frac{d^2P_6}{du^2}(u) \to -2u^3+\mathcal{O}(u^4)$ there are always stable circular orbits for large $r$, and because of $\frac{d^2P_6}{du^2}(\bQ^{-1}) = \frac{8(2-\bQ)}{\bQ^3}$ stable circular orbits near $r=0$ are possible if $\bQ>2$. The expression diverges at the zeros of $A(u)$, i.e.~only for $\bQ<\bQ_c$ at $a_{1,2}$. Also, $T(u)$ has double zeros at $\bQ_t\approx0.747$, $u_t\approx0.353$ and $\bQ_{\rm crit}$, $u_{\rm crit}$. Together with $T(b_1)>0$ for $\bQ>\bQ_{\rm crit}$ and $T(b_2)<0$ for $\bQ>2$ we infer:
\begin{itemize}
\item $0\leq\bQ\leq\bQ_{\rm crit}$: There is a single triple zero $u_{\rm ISCO}$ in $[0,a_1]$ which corresponds to the innermost stable circular orbit with stable circular orbits for $\br>\br_{\rm ISCO}:=(u_{\rm ISCO}^{-2}-\bQ^2)^{\frac{1}{2}}$ and unstable orbits for $\br<\br_{\rm ISCO}$.
\item $\bQ_{\rm crit}<\bQ\leq\bQ_c$: Here also a triple zero $u_{t_1}$ exists in $[0,a_1]$, but in addition $\frac{d^2P_6}{du^2}(u)<0$ for $u \in [a_2,b_1]$. Therefore, stable circular orbits exist in $\br>(u_{t_1}^{-2}-\bQ^2)^{\frac{1}{2}}$ and $(b_1^{-2}-\bQ^2)^{\frac{1}{2}} \leq \br < (a_2^{-2}-\bQ^2)^{\frac{1}{2}}$.
\item $\bQ_c<\bQ\leq\bQ_t$: Two triples zeros $u_{t_1}\leq u_{t_2}$ are located in $[0,b_1]$ with stable circular orbits for $\br>(u_{t_1}^{-2}-\bQ^2)^{\frac{1}{2}}$ and $(b_1^{-2}-\bQ^2)^{\frac{1}{2}} \leq \br < (u_{t_2}^{-2}-\bQ^2)^{\frac{1}{2}}$.
\item $\bQ_t<\bQ\leq2$: Here $u_{t_{1,2}}$ vanish and all possible circular orbits are stable, i.e.~for $\br \geq (b_1^{-2}-\bQ^2)^{\frac{1}{2}}$.
\item $2<\bQ$: A triple zero $u_{t_3}$ is located in the inner range $[b_2,\bQ^{-1}]$ and, therefore, stable circular orbits exist for $\br \geq (b_1^{-2}-\bQ^2)^{\frac{1}{2}}$ as before and $0 \leq \br < (u_{t_3}^{-2}-\bQ^2)^{\frac{1}{2}}$.
\end{itemize}
For a graphical representation of this analysis see Fig.~\ref{Fig:ISCO}, which also shows the radius of the innermost stable circular orbit as a function of $\bQ$. For the comparison with the circular orbits in Reissner-Nordstr\"om space-times, see Fig.~\ref{RNCircularOrbits}. There are more circular orbits in the Ay\'on-Beato--Garc\'\i{}a black hole space-time than in a Reissner-Nordstr\"om space-time.

\begin{figure}
\subfloat[(a) $\bQ=0.3$ ($\bQ<\bQ_{\rm crit}$)]{\includegraphics[width=0.42\textwidth]{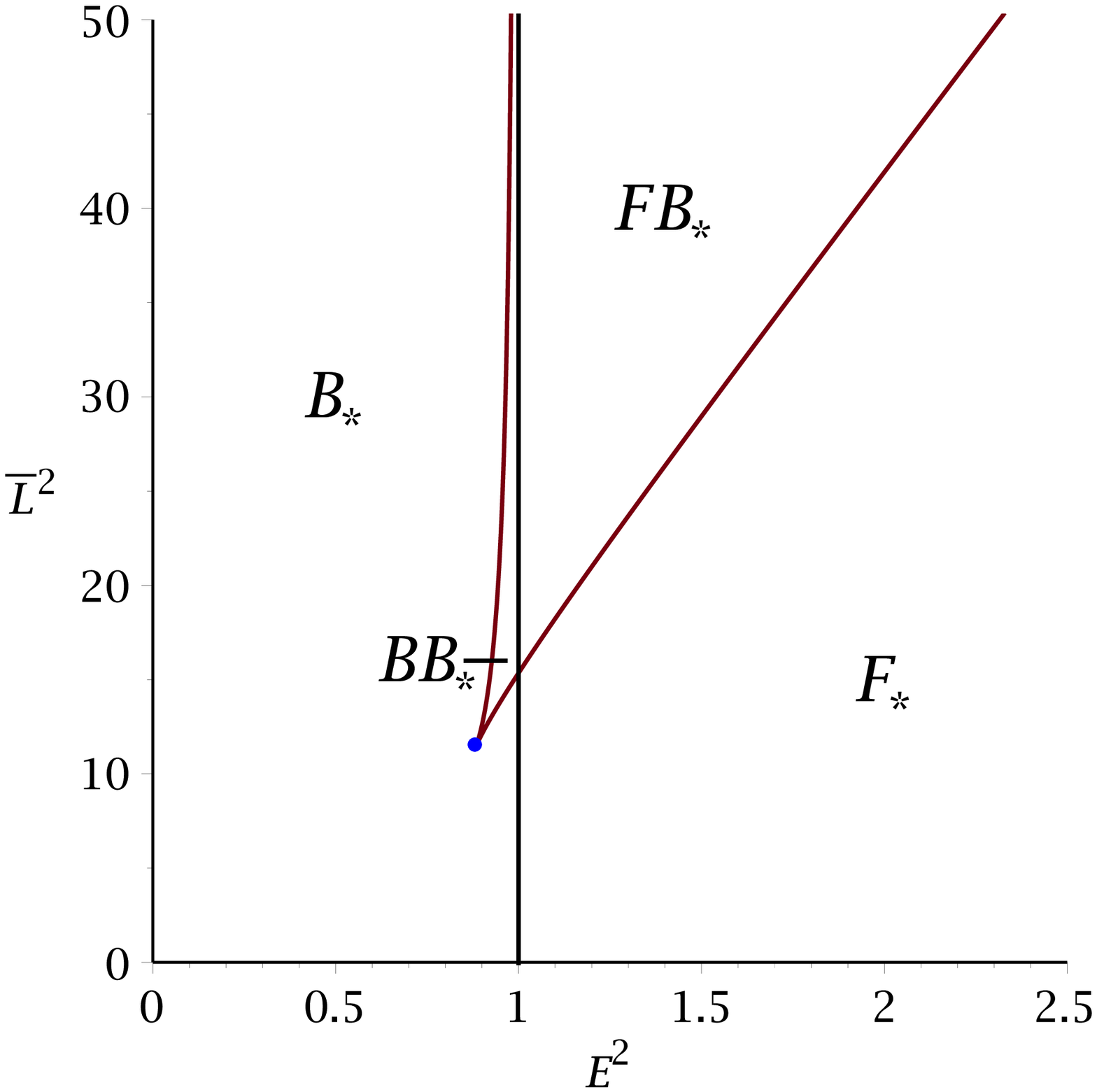}}\quad
\subfloat[(b) $\bQ=0.65$ ($\bQ_{\rm crit}<\bQ<\bQ_c$)]{\includegraphics[width=0.42\textwidth]{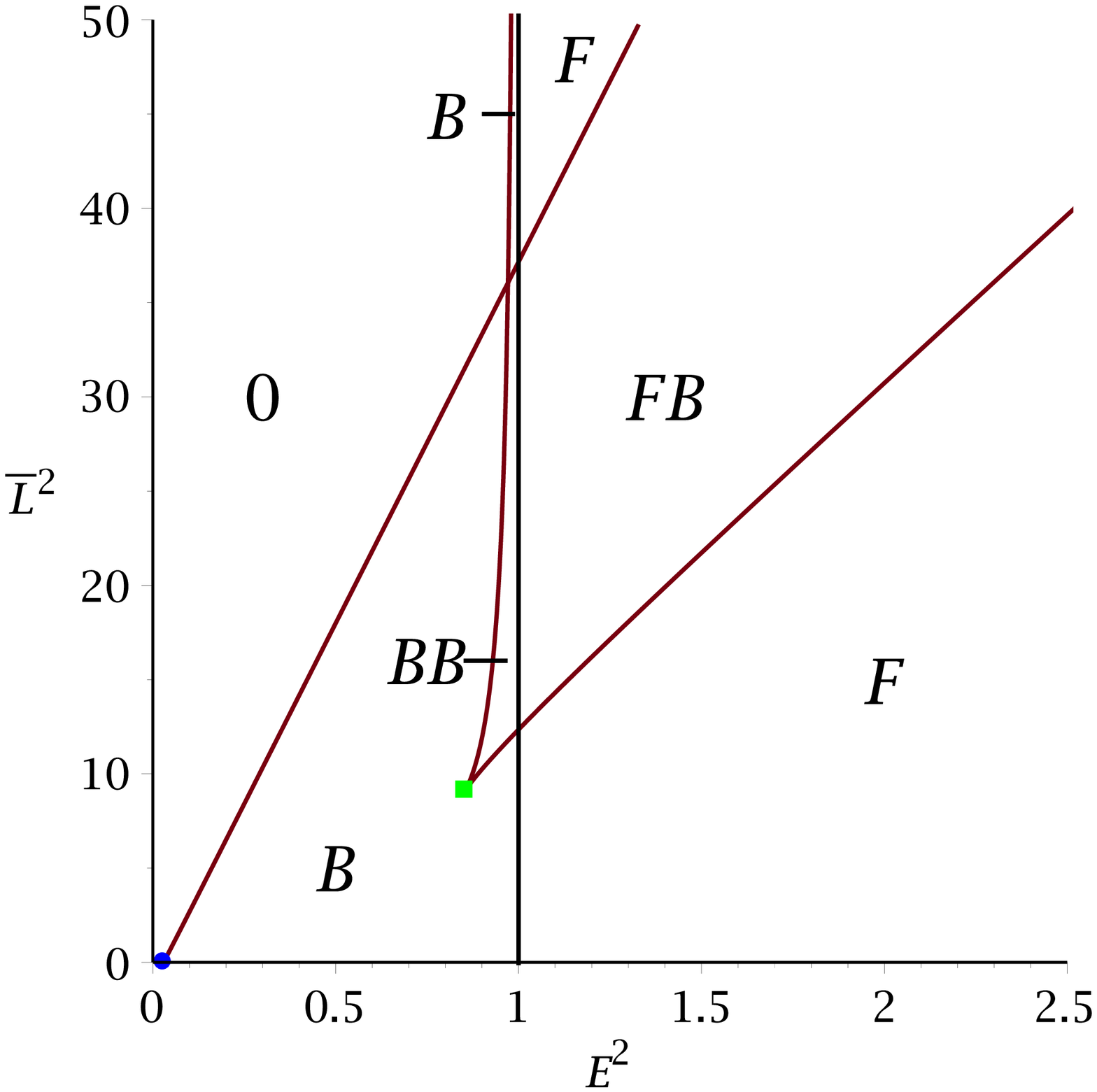}}\\
\subfloat[(c) $\bQ=0.7$ ($\bQ_{c}<\bQ<\bQ_t$)]{\includegraphics[width=0.42\textwidth]{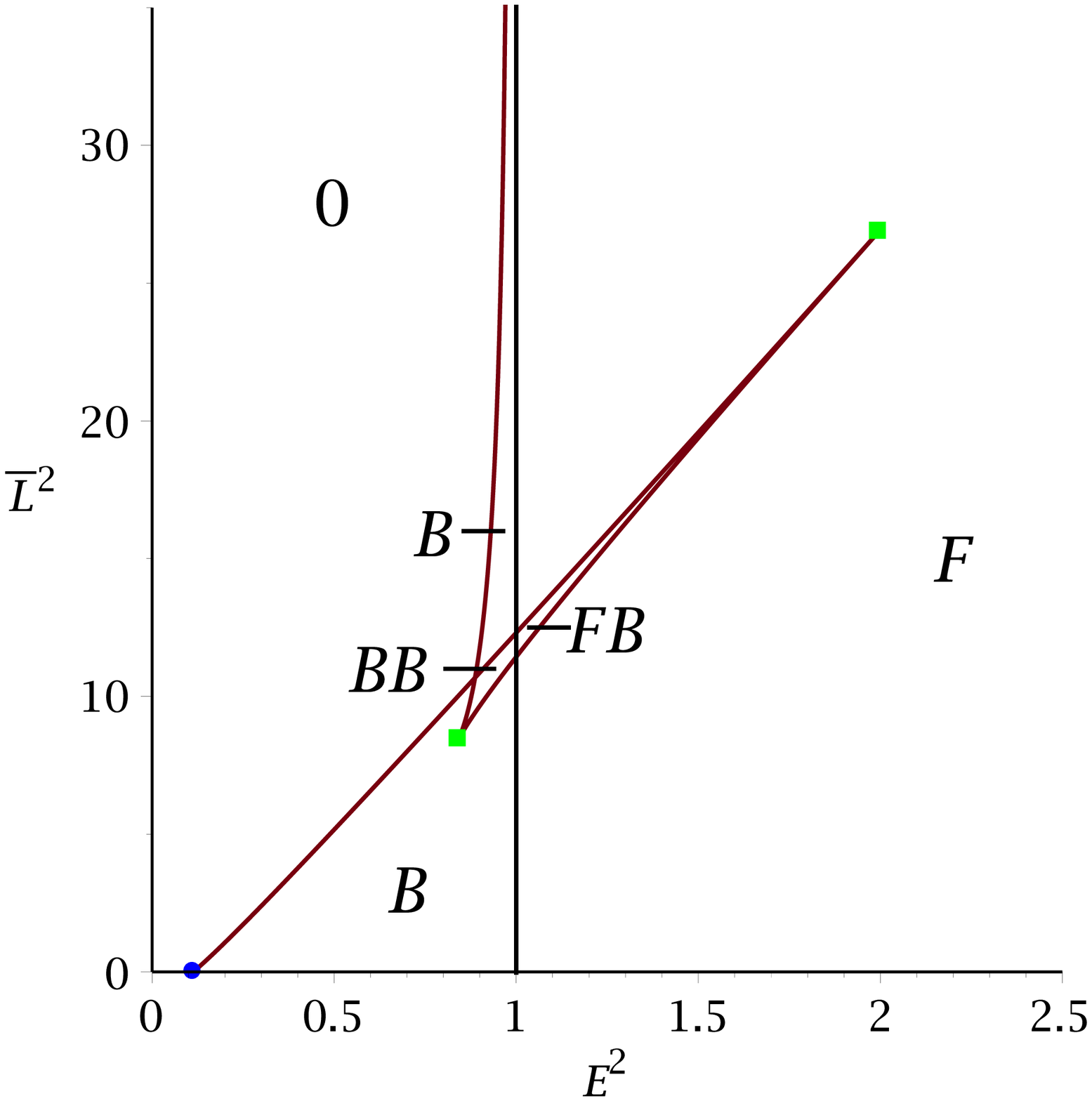}}\quad
\subfloat[(d) $\bQ=0.8$ ($\bQ_{t}<\bQ<2$)]{\includegraphics[width=0.42\textwidth]{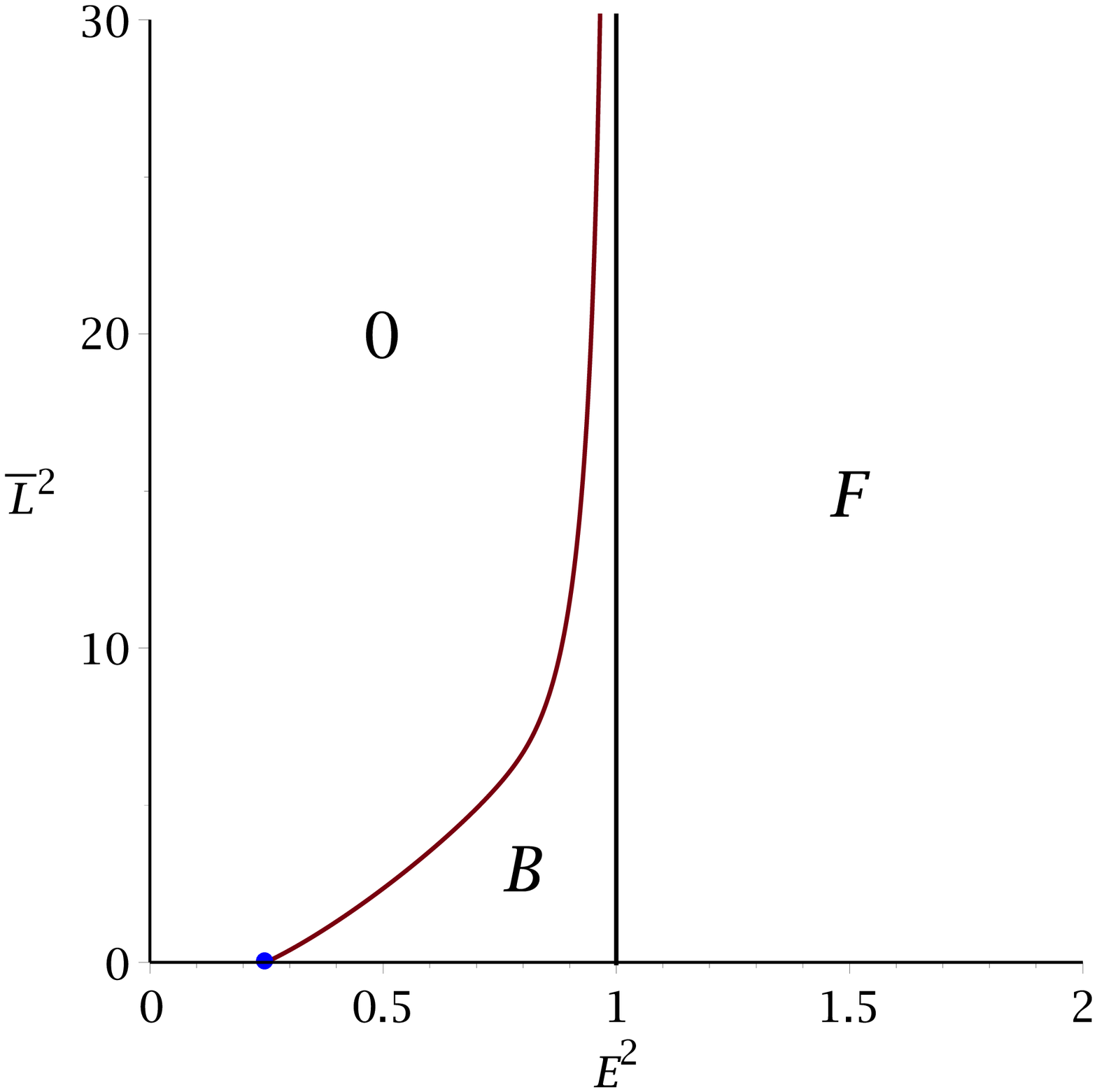}}\\
\subfloat[(e) $\bQ=5$ ($2<\bQ$)]{\includegraphics[width=0.42\textwidth]{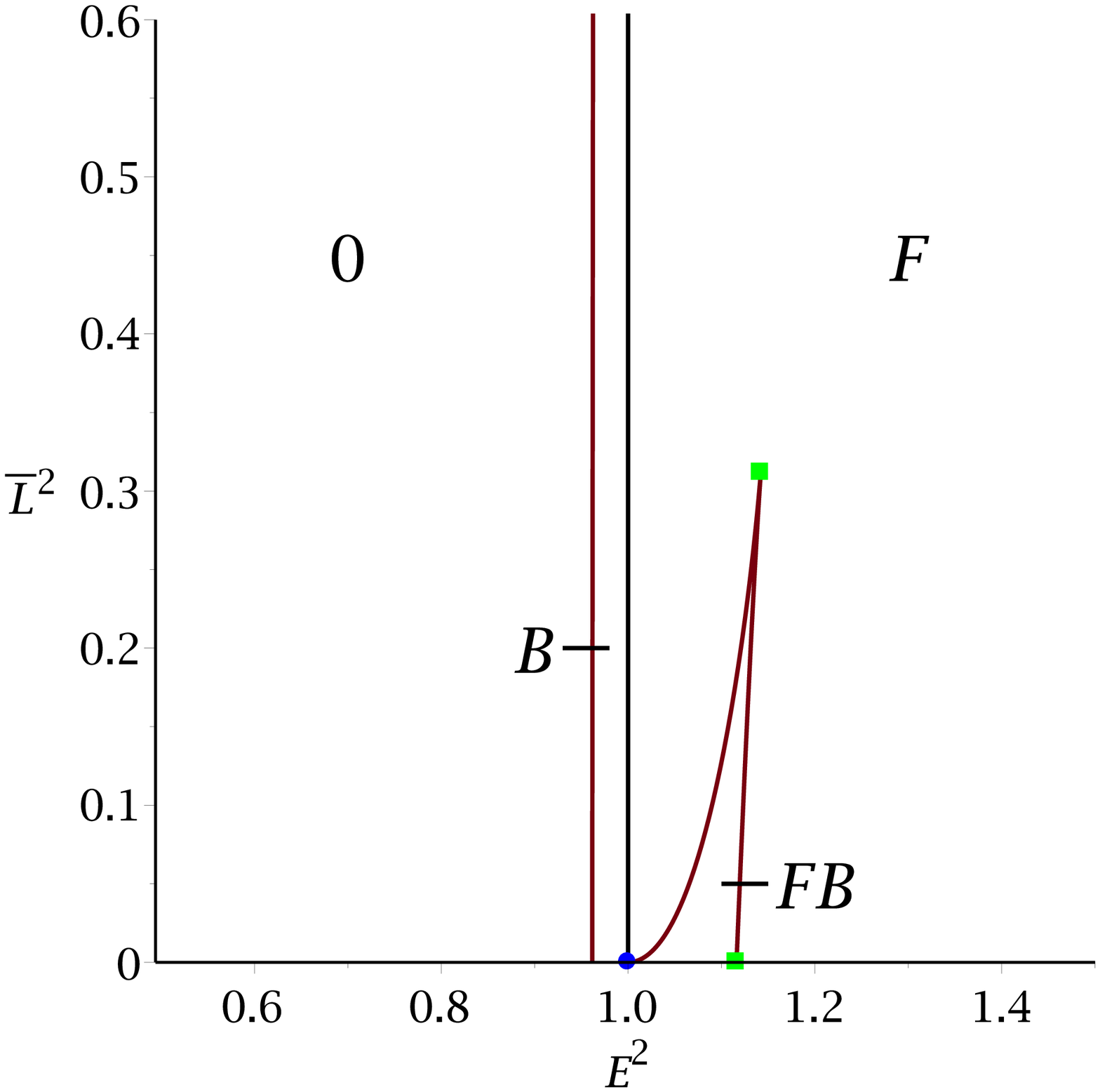}}
\caption{Orbit types for massive neutral test particles in Ay\'{o}n-Beato--Garc\'\i{}a space-times. The black vertical lines mark $E=1$, and the other red solid lines mark circular orbits. They divide different types of orbit configurations. Blue dots denote the innermost stable circular orbit $\br_{\rm ISCO}$ and green boxes the boundaries of regions of unstable circular orbits between $\br_{\rm ISCO}$ and $\br=\infty$. }
\label{Fig:Massive}
\end{figure}

With this analysis of circular orbits we can identify all possible orbits types in parameter space. In Fig.~\ref{Fig:Massive} the orbit types for massive neutral particles in Ay\'{o}n-Beato--Garc\'\i{}a space-times are shown. In regions marked with $F$ there is a flyby orbit ($r_{\rm p}\leq r \leq \infty$, $r_{\rm p}$ the periapsis), $B$ denotes a bound orbit ($r_{\rm p}\leq r\leq r_{\rm a}$, $r_{\rm a}$ the apoapsis), and an indexed star means that the orbit crosses both horizons. There are also regions where no motion is possible ($R(r)<0$ for all $r$) marked by $0$. If more than one type is possible the actual orbit is determined by the initial conditions.

As compared to the complete set of orbits of the Reissner-Nordstr\"om space-time (see Appendix) we find here a richer variety of orbits. For the black hole case, however, the structure of orbits is the same: we find (i) bound orbits crossing both horizons, (ii) two bound orbits where one crosses both horizons (the other then is like a standard planetary orbit), (iii) a flyby orbit crossing  both horizons, and (iv) a standard flyby orbit together with a bound orbit crossing the horizons. Only in the case of charges larger than the critical charge the manifold of orbits becomes richer and is different as can be seen from comparison of Fig.~\ref{Fig:Massive} with Fig.~\ref{Fig:RNMassive}. In particular, for large charges we here have bound orbits for energies $E>1$, and the energy interval below $E=1$ for which we have bound orbits is independent of $\bar L^2$. 

As a consequence, while for large charges we have particular types of orbits for orbital parameters which are not existent in the Reissner-Nordstr\"om case, for small charges for which we have black holes a difference between thee two types of charged black hole space-times can only be observed through the value of, e.g., the perihelion shift. This is what we will calculate later in this paper.

\subsection{Weakly charged test particles in black hole space-times}

From equation \eqref{chargedrds} it is a necessary condition for charged test particles that $R_{\bq}(\br)\geq0$ for a solution to exist. Let us first consider the cases $\br\to0,\infty$. At $\br\to\infty$ we have $R_{\bq}(\br)\to E^2-1-3\bq(E-\frac{3}{4}\bq)$ what implies that for $\frac{3}{2}\bq-1\leq E\leq \frac{3}{2}\bq+1$ infinity may not be reached. For $\br\to0$ the limit is not influenced by $\bq$ as $R_{\bq}(\br)\to -\frac{\bL^2}{\br^2}+E^2-1+\frac{\bL^2}{\bQ^3}(\bQ-2)$, i.e.~$\br=0$ may again only be reached for $\bL=0$ and $E^2\geq1$.

\begin{figure}
\centering
\includegraphics[width=0.5\textwidth]{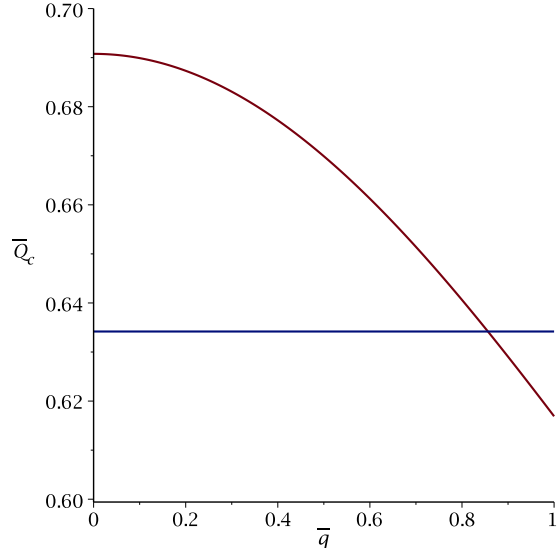}
\caption{The value $\bQ_{c}(\bq)$ as a function of the charge of the particle $\bq$. The blue horizontal line denotes the critical value $\bQ_{\rm crit}$.}
\label{Fig:chargeQc}
\end{figure}

\begin{figure}
\centering
\includegraphics[width=0.5\textwidth]{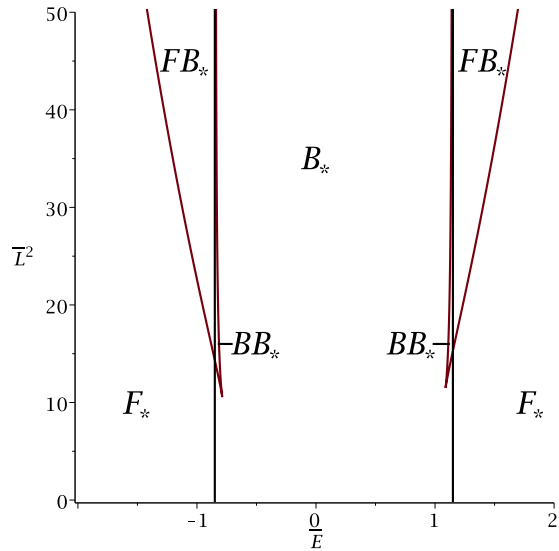}
\caption{Orbit types for charged test particles in Ay\'{o}n-Beato--Garc\'\i{}a space-times. Here $\bQ=0.4$ and $\bq=0.1$. The red lines mark circular orbits and the black lines $E=\frac{3}{2}\bq\pm1$. They divide different types of orbit configurations, which are denoted in the same way as for neutral test particles.}
\label{Fig:chargeEL}
\end{figure}

We now turn to the simpler equation \eqref{chargeduds}. It is necessary that $U(u)\geq0$ for the existence of a solution. This means that the type of orbit is determined by the values of $U$ at the boundaries of the physical meaningful region $u \in [0,\bQ^{-1}]$ and the number of real zeros in between. As noted above, the sign of $U$ at $u=0$ changes at $E=\frac{3}{2}\bq\pm1$ whereas $U$ is negative at $u=\bQ^{-1}$ as long as $\bL\neq0$. The number of real zeros of $U$ in $[0,\bQ^{-1}]$ changes for varying constants of motion if double zeros occur, which correspond to circular orbits. Solving $U(u)=0$ and $\frac{dU}{du}(u)=0$ for $E$ and $\bL^2$ yields
\begin{equation}\label{chargeEL}
\begin{aligned}
E_{1,2} & = \frac{1}{4A(u)}\big[\bq (1-\bQ^2u^2)^{\frac52} D(u) \pm g_{tt}(u)\sqrt{16A(u)+\bq^2C(u)}\big] \\
\bL^2_{1,2} & = \frac{(1-\bQ^2u^2)^2B(u)}{uA(u)} + \frac{\bq\bQ^2g_{tt}(u)}{8uA^2(u)} (12\bQ^2u^2-15u-2) \\
& \quad \times \big[ \bq \bQ^2(1-\bQ^2u^2)^6(12\bQ^2u^2-15u-2)u \mp (1-\bQ^2u^2)^{\frac72}\sqrt{16A(u)+\bq^2C(u)} \big]
\end{aligned}
\end{equation}
where $g_{tt}(u)$ is the metric function under the substitution $u=1/\sqrt{\br^2+\bQ^2}$, $C(u)=u^2\bQ^4(1-\bQ^2u^2)^5(12\bQ^2u^2-15u-2)^2$ and
\begin{align}
D(u) & = 4\bQ^8u^7-15\bQ^6u^6+2\bQ^4(6+\bQ^2)u^5-5\bQ^4u^4-6\bQ^2(3\bQ^2-1)u^3\nonumber\\
& \quad +35\bQ^2u^2-2(9+\bQ^2)u+6\,.
\end{align}
For $E$ and $\bL$ to be real the expression under the square root has to be positive, $16A(u)+\bq^2C(u)\geq0$. From the discussion of neutral particles we know that $A(u)>0$ for all $u \in [0,\bQ^{-1}]$ if $\bQ>\bQ_c$. In addition, $C(u)\geq0$ for all $u$ in the admissible range. Let us denote the smallest charge $\bQ$ for which $16A(u)+\bq^2C(u)\geq0$ for all $u \in [0,\bQ^{-1}]$ by $\bQ_{c}(\bq)$ (i.e.~$\bQ_c(\bq=0)=\bQ_c$). The value of $\bQ_{c}(\bq)$ as a function of $\bq$ is shown in Figure \ref{Fig:chargeQc}. Above $\bQ_{c}(\bq)$ the expressions for $E_{1,2}$ and $\bL^2_{1,2}$ from \eqref{chargeEL} are real for all $u \in [0,\bQ^{-1}]$. Below it $16A(u)+\bq^2C(u)$ has two zeros $\tilde{a}_{1,2}$ with $16A(u)+\bq^2C(u)\geq0$ for $0\leq u\leq\tilde{a}_1$ and $\tilde{a}_2\leq u\leq\bQ^{-1}$. As a second necessary condition, $\bL^2$ has also to be greater or equal to zero. At the boundaries we have $\bL_{1,2}^2 = (1\pm \bq\bQ^2)u^{-1} + \mathcal{O}(u^0)$ and $\bL_{1,2}^2 = 4\bQ^3(\bQ-2)(u-\bQ^{-1})^2 + \mathcal{O}((u-\bQ^{-1})^3)$. This means that near $u=0$ ($r=\infty$), $\bL_1^2$ is always positive but $\bL_2^2$ is positive only if $\bq\bQ^2<1$. Near $u=\bQ^{-1}$ ($r=0$), $\bL_{1,2}^2$ is negative if $\bQ<2$. From the discussion of neutral test particles we know that for $\bq=0$ the expression $\bL^2$ is positive in the range $[0,a_1]$. This remains valid for small $\bq$ until that value of $\bq(\bQ)$ is reached where $\bL_{2}^2$ has a double zero in $[0,a_1]$. We therefore assume here that the test-particle is weakly charged in the sense that $\bq\bQ^2<1$ and $\bq$ that small that $\bL_2^2$ is positive in the range $[0,\tilde{a}_1]$. Numerical analysis indicates that $\bq<\bQ_{\rm crit}$ seems to be sufficient. Note that in this case always $\bQ_c(\bq)>\bQ_{\rm crit}$. A typical example of orbit configurations for fixed $\bQ$ and $\bq$ is shown in Figure \ref{Fig:chargeEL}.

\section{Analytical solution for motion of neutral test particles}

In this section we derive the analytical solution to the equation of motion \eqref{drdphi} in the Ay\'{o}n-Beato--Garc\'\i{}a space-time. With the dimensionless quantities used throughout the paper and the substitution $u=1/\sqrt{\br^2+\bQ^2}$ this equation reads
\begin{align}
\left(\frac{du}{d\varphi}\right)^2 & = \frac{(1-\bQ^2u^2)^2}{L^2} P_6(u) \label{dudphi}
\end{align}
This ordinary differential equation can be solved in terms of algebro-geometric methods. It corresponds to a hyperelliptic curve of genus two. This situation is similar to the structure of the geodesic equation in Schwarzschild-de Sitter space-time, where an analytical solution can be found in terms of derivatives of the Riemann $\theta$-function in two complex variables restricted to the theta-divisor, see \cite{HackmannLaemmerzahl08}. However, here the differential is of the third kind, which introduces an additional complication, which can be handled in the following way: by introducing a parameter $\lambda$ with $\frac{d\lambda}{d\bar s} =u^2$, $\lambda({\bar s}_0)=0$ in equation \eqref{dudphi} we split the problem in two parts,
\begin{align}
\left(\frac{du}{d\lambda}\right)^2 & = P_6(u)\,, \label{dudlambda}\\
\frac{d\varphi}{d\lambda} & = \frac{\bL}{1-\bQ^2u^2}\,, \label{dphidlambda}
\end{align}
with $u(0)=u_0$, $\varphi(0)=\varphi_0$. This new affine parameter $\lambda$ can be seen as an analog of the Mino time \cite{Mino03}. Let us first consider \eqref{dudlambda}. With the substitution $u=\pm\frac{1}{x}+u_P$ where $P_6(u_P)=0$ the problem is transformed to the standard form
\begin{align}
\left(x \frac{dx}{d\lambda}\right)^2 & = c_5 \sum_{i=0}^5 \frac{c_i}{c_5} x^i = c_5 P_5(x)\,, \quad c_{i} = \frac{(\pm 1)^i}{(6-i)!} \frac{d^{(6-i)} P_{6}}{du^{(6-i)}} (u_P)\,.\label{dxdlambda}
\end{align}
Here, the sign in the substitution should be chosen such that $c_5$ is positive and, therefore, depends on the choice of $u_P$. The solution to this equation is given by \cite{BuchstaberEnolskiiLeykin97,EnolskiiPronineRichter03,HackmannLaemmerzahl08}
\begin{align}
x(\lambda) = - \frac{\sigma_1}{\sigma_2}(\lambda_\sigma)\,,\label{solx}
\end{align}
where $\sigma_i$ is the i-th derivative of the Kleinian sigma function in two variables
\begin{align}
\sigma(z) = C e^{z^t\kappa z} \theta[K_\infty]((2\omega)^{-1}z;\tau)
\end{align}
which is given by the Riemann $\theta$-function with characteristic $K_\infty$. A number of parameters enters here: the symmetric Riemann matrix $\tau$, the period-matrix $(2\omega,2\omega')$, the period-matrix of the second kind $(2\eta,2\eta')$, the matrix $\kappa=\eta(2\omega)^{-1}$, and the vector of Riemann constants with base point at infinity $2 K_\infty=(0,1)^t+(1,1)^t\tau$. The constant $C$ can be given explicitly, see e.g.~\cite{BuchstaberEnolskiiLeykin97}, but does not matter here. In eq.~\eqref{solx} the argument $\lambda_\sigma$ is an element of the one-dimensional sigma divisor: $\lambda_\sigma=(f(\sqrt{c_5}\lambda-\lambda_{\rm in}),\sqrt{c_5}\lambda-\lambda_{\rm in})^t$ where $\lambda_{\rm in} = \int_{x_0}^{\infty} \frac{xdx}{\sqrt{P_5(x)}}$ with $x_0=\pm(u_0-u_P)^{-1}$ depends only on the initial values and the function $f$ is given by the condition $\sigma(\lambda_\sigma)=0$. For more details on the construction of such solutions see e.g.~\cite{BuchstaberEnolskiiLeykin97}. With \eqref{solx} the solution for $\br$ is given by
\begin{align}
\br^2(\lambda)= \frac{\sigma_1^2(\lambda_\sigma)}{(\sigma_2(\lambda_\sigma)\mp u_P\sigma_1(\lambda_\sigma))^2} - \bQ^2\,,\quad P_6(u_P)=0\,, \,\sigma(\lambda_\sigma)=0\,. 
\label{solr}
\end{align}

\begin{figure}
\centering
\includegraphics[width=0.4\textwidth]{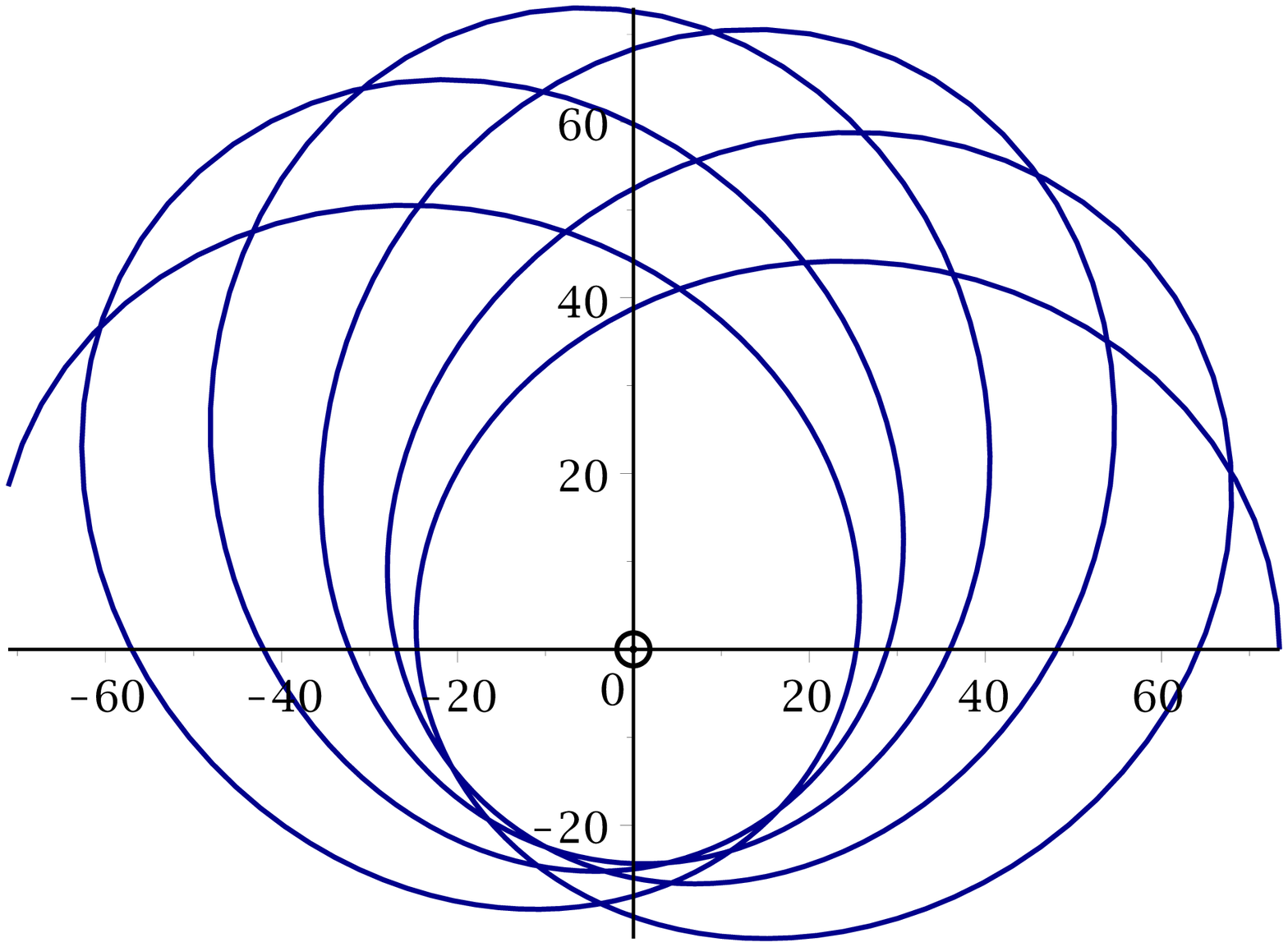}\quad
\includegraphics[width=0.4\textwidth]{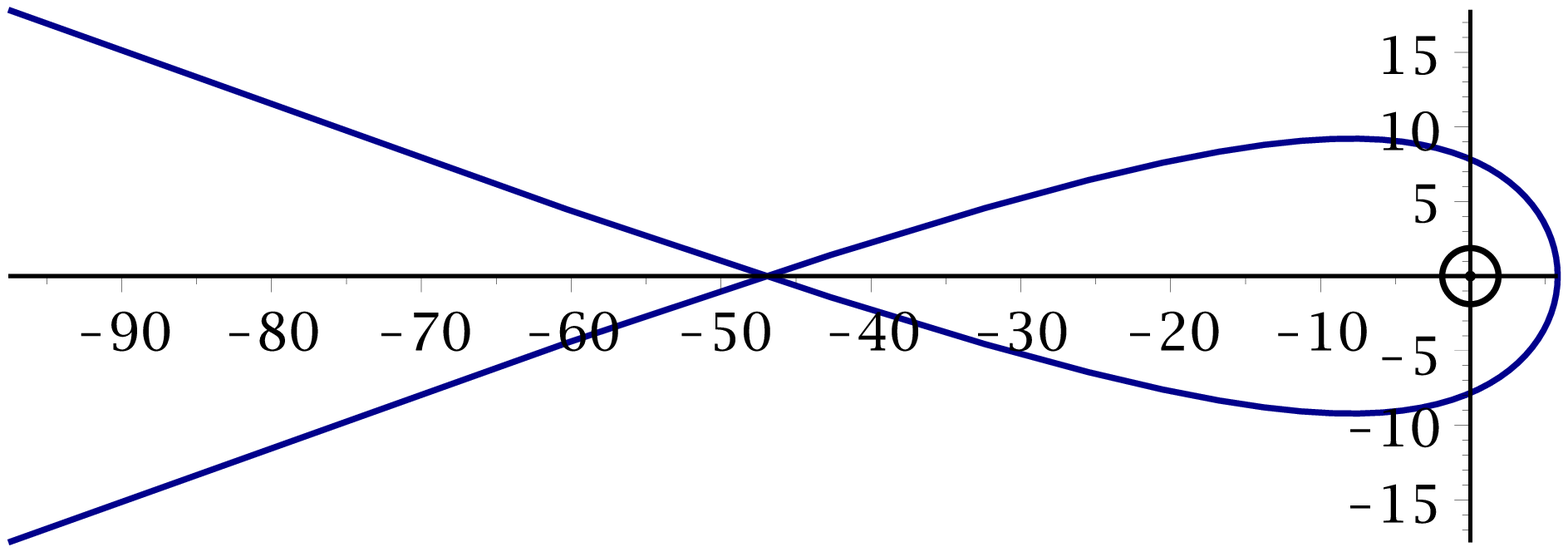}
\caption{Orbits for neutral massive particles in a space-time with $\bQ=0.3$. Left: bound orbit for $E^2=0.98$, $\bL^2=40$. Right: flyby orbit for $E^2=1.05$, $L^2=20$. The small black circles indicate the horizons.}
\label{Fig:orbits}
\end{figure}

Let us now turn to the equation \eqref{dphidlambda} for $\varphi$. It can be written in the form
\begin{align}
\varphi-\varphi_0 & = \int_{\lambda_0}^\lambda \frac{\bL d\lambda}{1-\bQ^2u^2} = \frac{\bL}{\sqrt{c_5}} \int_{x_0}^x \frac{xdx}{(1-\bQ^2u^2(x))\sqrt{P_5(x)}}\,,\\
& = \frac{\bL}{\sqrt{c_5}} \Bigg[ \int_{x_0}^x \frac{2\bQ^2u_P dx}{(1-\bQ^2u_P^2)^2 \sqrt{P_5(x)}} + \int_{x_0}^x \frac{xdx}{(1-\bQ^2u_P^2)\sqrt{P_5(x)}}\nonumber\\
& \qquad + \sum_{i=1}^2 \int_{x_0}^x \frac{C_idx}{(x-u_i)\sqrt{P_5(x)}} \Bigg]\,,
\end{align}
where $u_{1,2}=\frac{\mp\bQ}{1\pm\bQ u_P}$ and $C_i=(-1)^i\frac{u_i^3}{2\bQ}$. The first two terms can be expressed directly in terms of $\lambda$,
\begin{align}
\int_{x_0}^x \frac{xdx}{\sqrt{P_5(x)}} & = \sqrt{c_5}\lambda = (\lambda_\sigma-\lambda_{\sigma,\lambda=0})_2\,,\\
\int_{x_0}^x \frac{dx}{\sqrt{P_5(x)}} & = \int_{x_0}^\infty \frac{dx}{\sqrt{P_5(x)}} + \int_\infty^{x} \frac{dx}{\sqrt{P_5(x)}}\nonumber \\
& = - f(-\lambda_{\rm in}) + f(\sqrt{c_5}\lambda-\lambda_{\rm in}) = (\lambda_\sigma-\lambda_{\sigma,\lambda=0})_1\,.
\end{align}
The summands of the last term can be rewritten as \cite{BuchstaberEnolskiiLeykin97}
\begin{align}
\int_{x_0}^x \frac{dx}{(x-u_i)\sqrt{P_5(x)}} & = \frac{1}{\sqrt{P_5(u_i)}} \bigg[ \frac{1}{2} \log \frac{\sigma(\Sigma^+(\lambda))}{\sigma( \Sigma^{-}(\lambda) )} - \frac{1}{2} \log \frac{\sigma( \Sigma^{+}(0))}{\sigma( \Sigma^{-}(0) )} \nonumber \\
& \quad - (\lambda_\sigma-\lambda_{\sigma,\lambda=0})^t \left( \int_{u_i^-}^{u_i^+} dr_j\right)_{j=1,2} \bigg]
\end{align}
where $\Sigma^{\pm}(\lambda)_j = (\lambda_\sigma)_j - 2\int_\infty^{u_i^{\pm}} \frac{x^{j-1}dx}{\sqrt{P_5(x)}}$ and $dr_j=\sum_{k=j}^{5-j} (k+1-j) \frac{c_{k+1+j}}{c_5} \frac{x^k dx}{4 \sqrt{P_5(x)}}$ with $c_i$ as in \eqref{dxdlambda}. Here the sign in $u_i^{\pm}$ indicates the branch of the square root. This means, $\int_{u_i^-}^{u_i^+} dr_j = \pm 2 \int_{u_i}^{e_i} dr_j$ (modulo periods) where $e_i$ is a zero of $P_5$ close to $u_i$ . Summarized this gives
\begin{align}
\varphi(\lambda) & = \frac{\bL}{\sqrt{c_5}} \Bigg\{ \frac{2\bQ^2u_P(f(\sqrt{c_5}\lambda-\lambda_{\rm in})- f(-\lambda_{\rm in}))}{(1-\bQ^2u_P^2)^2}  + \frac{\sqrt{c_5}\lambda}{1-\bQ^2u_P^2} + \sum_{i=1}^2 \frac{C_i}{\sqrt{P_5(u_i)}} \times \nonumber\\
& \quad \times \bigg[ \frac{1}{2} \log \frac{\sigma(\Sigma^+(\lambda))}{\sigma( \Sigma^{-}(\lambda) )} - \frac{1}{2} \log \frac{\sigma( \Sigma^{+}(0))}{\sigma( \Sigma^{-}(0) )} - (\lambda_\sigma-\lambda_{\sigma,\lambda=0})^t \left( \int_{u_i^-}^{u_i^+} dr_j\right)_{j=1,2} \bigg] \Bigg\}\,. \label{solphi}
\end{align}
The equations \eqref{solr} and \eqref{solphi} together analytically solve the differential equation \eqref{drdphi} in a parametric form. In Fig.~\ref{Fig:orbits} we used this parametric solution to plot as an example two orbits of neutral test particles in the Ay\'{o}n-Beato--Garc\'\i{}a space-time, which do not cross the horizons.

\section{Periastron shift for neutral test particles}

The equation \eqref{drdphi} for neutral particles can be used to derive the periastron shift of bound orbits in Ay\'{o}n-Beato--Garc\'\i{}a space-times, that is, the difference between $2\pi$ and the azimuthal angle accumulated from one passage of the periastron to the next. If we introduce the normalized quantities used throughout the paper eq.~\eqref{drdphi} can be rewritten as
\begin{align}
\left(\frac{d\br}{d\varphi}\right)^2 & =  \frac{\br^4}{\bL^2} \left[E^2 - \left( 1- \frac{2\br}{(\br^2+\bQ^2)^\frac32} + \frac{\bQ^2\br^2}{(\br^2+\bQ^2)^2}\right) \left(1 + \frac{L^2}{r^2}\right)\right] =: R_\varphi(\br)\,.
\end{align}
The periastron shift $\Omega_r$ is then given by the period $\Lambda_r$ of $r(\varphi)$, i.e.~$r(\varphi+\varLambda_r)=r(\varphi)$, minus $2 \pi$, which is
\begin{align}
\Omega_r = \Lambda_r -2\pi = 2 \int_{\br_{\rm p}}^{\br_{\rm a}} \frac{d\br}{\sqrt{R_\varphi(\br)}} - 2\pi\,, \label{periastron}
\end{align}
where $\br_{\rm p}$ is the periastron and $\br_{\rm a}$ the apastron.

As the Ay\'{o}n-Beato--Garc\'\i{}a space-time approaches the Reissner-Nordstr\"om space-time for large radii and the Schwarzschild space-time for small $\bQ$, the question arises how the expression \eqref{periastron} differs from the analogous expression in these space-times. To analyze this, we consider how $\Omega_r$ expands for small charges $\bQ$ and then compare the result to the Reissner-Nordstr\"om correction for small charges. For this we consider $\br_{\rm p}$ and $\br_{\rm a}$ as fixed and all other quantities like $E=E(\bQ)$ and $\bL=\bL(\bQ)$ as dependent on $\bQ$. This means that the derivatives of $E$ and $\bL$ with respect to $\bQ$ are needed. Therefore, we consider the zeros of $R_\varphi(\br)$ which are given by $\pm \sqrt{\br_i^2-\bQ^2}$ where $\br_i$ are the six zeros of
\begin{align}
P(\br) & = (E^2-1)\br^6+2\br^5-(E^2\bQ^2+\bL^2)\br^4+2(\bL^2-2\bQ^2)\br^3+\bQ^2(2\bQ^2-\bL^2)\br^2\nonumber\\
& \quad +2\bQ^2(\bQ^2-\bL^2)\br-\bQ^6+\bQ^4\bL^2\,. \label{defri}
\end{align}
As $\br_{\rm p}$ and $\br_{\rm a}$ should be fixed this implies
\begin{align}
\br_{1,2} & =\sqrt{\br_{\rm p,a}^2+\bQ^2}\approx \br_{\rm p,a} + \frac{\bQ^2}{2\br_{\rm p,a}}\,.
\end{align}
By expanding $P(\br)=(E^2-1)\prod_{i=1}^6(\br-\br_i)$ in powers of $\bQ$ and comparing coefficients we can then derive that
\begin{align}
E & \approx \left(\frac{(\br_{\rm a}-2)(\br_{\rm p}-2)(\br_{\rm a}+\br_{\rm p})}{\br_{\rm a}\br_{\rm p}(\br_{\rm a}+\br_{\rm p}-2)-2(\br_{\rm a}^2+\br_{\rm p}^2)}\right)^\frac12 \nonumber\\
& \quad  - \frac{(\br_{\rm a}+\br_{\rm p})(7\br_{\rm a}\br_{\rm p}-4\br_{\rm a}-4\br_{\rm p}-12)}{2((\br_{\rm a}-2)(\br_{\rm p}-2)(\br_{\rm a}+\br_{\rm p}))^\frac12 (\br_{\rm a}\br_{\rm p}(\br_{\rm a}+\br_{\rm p}-2)-2(\br_{\rm a}^2+\br_{\rm p}^2))^\frac32}  \bQ^2\,,\\
\bL & \approx \left(\frac{2}{\br_{\rm a}\br_{\rm p}(\br_{\rm a}+\br_{\rm p}-2)-2(\br_{\rm a}^2+\br_{\rm p}^2)}\right)^\frac12 \nonumber\\
& \quad  - \frac{(\br_{\rm a}+\br_{\rm p})(\br_{\rm a}^2\br_{\rm p}+\br_{\rm a}\br_{\rm p}^2+\br_{\rm a}\br_{\rm p}+3\br_{\rm a}^2+3\br_{\rm p}^2-6\br_{\rm a}-6\br_{\rm p})}{2\sqrt{2}(\br_{\rm a}\br_{\rm p}(\br_{\rm a}+\br_{\rm p}-2)-2(\br_{\rm a}^2+\br_{\rm p}^2))^\frac32}\bQ^2\,.
\end{align}
Now the Taylor expansion of \eqref{periastron} reads
\begin{align}
\Omega_r & \approx \Lambda_{r,S} + \Lambda_{r,Q^2}\bQ^2  -2\pi \nonumber\\
& = 2 \int_{\br_{\rm p}}^{\br_{\rm a}} \frac{d\br}{\sqrt{R_0(\br)}}  -2\pi \nonumber\\
& \quad + \int_{\br_{\rm p}}^{\br_{\rm a}} \frac{(\bL''E^2-\bL E'' E-\bL'')\br^5+2\br^4\bL''+\bL\br^3+3\bL\br^2+\bL^3\br+3\bL^3)|_{\bQ=0}d\br}{\bL^3\br R_0(\br)\sqrt{R_0(\br)}} \bQ^2,\label{seriesperi}
\end{align}
where $R_0(\br)=(E^2-1)\bL^{-2}\br^4+2\bL^{-2}\br^3-\br^2+2\br$ is the Schwarzschild expression and a prime denotes differentiation with respect to $\bQ$. Accordingly, the first term of the expansion yields the Schwarzschild periastron precession rate, as can be seen by substituting $\br=\frac{\alpha nx^2+\beta}{nx^2+1}$ with $\alpha=\br_{3,0}$, $\beta=\br_{\rm p}$, and $n=\frac{\br_{\rm a}-\br_{\rm p}}{\br_{3,0}-\br_{\rm a}}$, where $0<\br_{3,0}<\br_{\rm p}<\br_{\rm a}$ are the zeros of $R_0(\br)$. The first term in \eqref{seriesperi} is then given by
\begin{align}
\Lambda_{r,{\rm S}} & = \frac{4L(0)}{\sqrt{(1-E^2(0))\br_{\rm p}(\br_{\rm a}-\br_{3,0})}} \int_0^1 \frac{dx}{\sqrt{(1-x^2)(1-k^2x^2)}}\nonumber\\
& = \frac{4\sqrt{\br_{\rm a}\br_{\rm p}} K(k)}{\sqrt{\br_{\rm a}\br_{\rm p}-4\br_{\rm a}-2\br_{\rm p}}} = \frac{4\sqrt{p}K(k)}{\sqrt{p+2e-6}}\,,
\end{align}
where $K(k)$ is the complete elliptic integral of the first kind with the modulus $k^2=\frac{2(\br_{\rm a}-\br_{\rm p})}{\br_{\rm a}\br_{\rm a}-4\br_{\rm p}-2\br_{\rm a}} = \frac{4e}{p+2e-6}$. Here $\br_{\rm a}=\frac{p}{1-e}$ and $\br_{\rm p}=\frac{p}{1+e}$ with the semilatus rectum $p$ and the eccentricity $e$. This is the standard result. By applying the same substitution to the second term in \eqref{seriesperi} and a decomposition in partial fractions we get
\begin{align}
\Lambda_{r,Q^2} = \frac{(3p^2-8p+e^2+3)K(k)}{p^\frac32\sqrt{p+2e-6}} + \frac{(3p^3-24p^2+75p-7pe^2-12(1-e^2))E(k)}{p^\frac32\sqrt{p+2e-6}\,(2e-p+6)}\,, \label{ABGPerihel}
\end{align}
where $E(k)$ is the complete elliptic integral of the second kind.

A corresponding analysis of the periapsis shift in Reissner-Nordstr\"om space-time yields
\begin{align}
\Omega_{r} & \approx \Lambda_{r,{\rm S}} -2\pi + \frac{\bQ^2}{\sqrt{p}\sqrt{p+2e-6}} \left[ (p-2)K(k) - \frac{(p^2-6p-2e^2+18)E(k)}{p-2e-6} \right] \label{RNPerihel}
\end{align}
Obviously the two expressions for the periastron shift differ in the strong field, but for $p\to\infty$ we obtain in both cases $\Omega_{r} \approx \frac{6\pi}{p} - \frac{\pi}{p} \bQ^2$. This result coincides with the result of Chaliasos \cite{Chaliasos2001} (see his equation (47) with vanishing charge of the test particle, i.e.~$e=0$). The expressions \eqref{ABGPerihel} and \eqref{RNPerihel} will then serve as basis for a future comparison with observational data and subsequent analysis whether a singular or regular black hole is responsible for the motion of objects orbiting the black hole.

\section{Summary}

In this paper we considered the motion of massive test particles in the metric presented in \cite{AyonBeatoGarcia98}, which we called the Ay\'{o}n-Beato--Garc\'\i{}a space-time. It is given as a solution to the Einstein equations coupled to a nonlinear electrodynamics, and is completely determined by its mass $M$ and its charge $Q$. After a review of the space-time and the corresponding equations of motion we classified the complete set of orbit types for neutral test particles moving on geodesics, without any restriction on the value of $Q$. In particular, we analyzed conditions for circular orbits and the position of the innermost stable circular orbit as a function of the charge. We also considered possible types of orbits of a weakly charged test particle moving in a black hole space-time. In addition, we derived the analytical solution of the equation of motion dependent on a new affine parameter, which can be seen as an analog of the Mino time \cite{Mino03}. We also discussed the periastron precession rate and derived a post-Schwarzschild correction to the order $\bQ^2$. A more detailed analysis of the comparison with possible astronomical observations is postponed to future work. It would also be interesting to extend this work to a rotating version of the Ay\'{o}n-Beato--Garc\'\i{}a solution, which however still needs to be derived.

\section*{Acknowledgments}
We thank Volker Perlick for useful discussions. This research was supported by the DFG Research Training Group 1620 “Models of Gravity”, by DFG–CONACyT Grant No.~B330/418/11 -- No. 211183, by GIF Grant No.~1078/2009, and by CONACyT Grants No.~166041F3 and No.~178346F3, as well as from FP7, Marie Curie Actions, People (IRSES-606096) .


\section*{Appendix: Orbits in Reissner-Nordtstr\"om space-times}

\begin{figure}[th!]
\subfloat[(a) $\bar Q = 0$: Schwarzschild]{\includegraphics[width=0.42\textwidth]{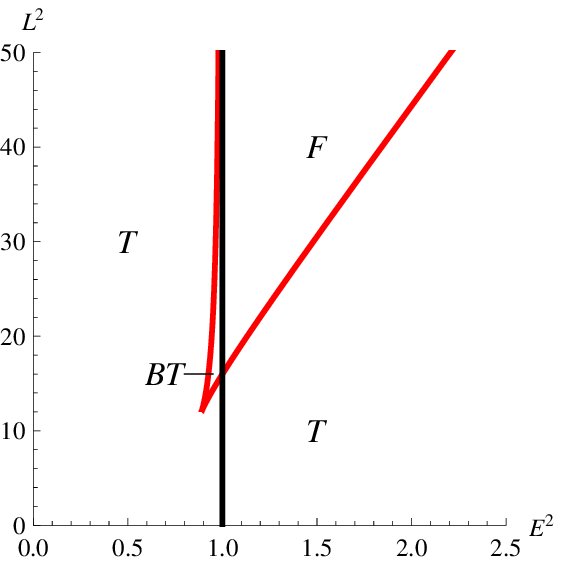}}\quad
\subfloat[(b) $\bar Q = 0.5$]{\includegraphics[width=0.42\textwidth]{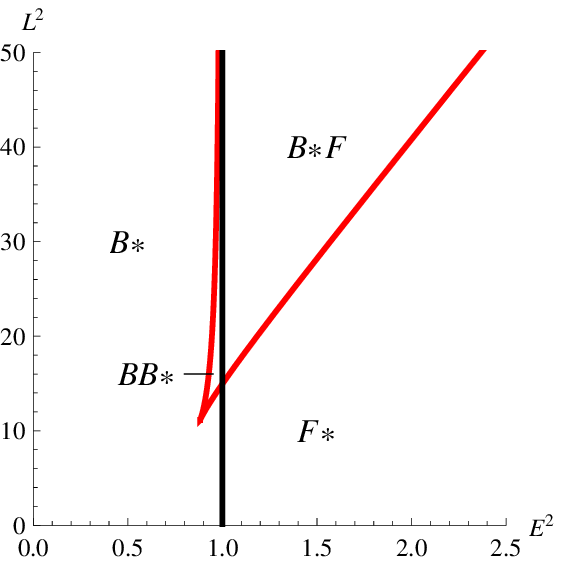}}\\
\subfloat[(c) $\bar Q = 1$: extremal Reissner-Nordstr\"om]{\includegraphics[width=0.42\textwidth]{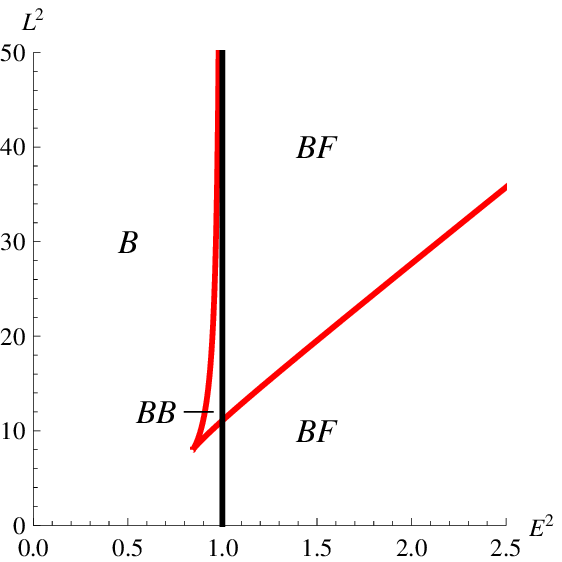}}\quad
\subfloat[(d) $\bar Q = 1.07$]{\includegraphics[width=0.42\textwidth]{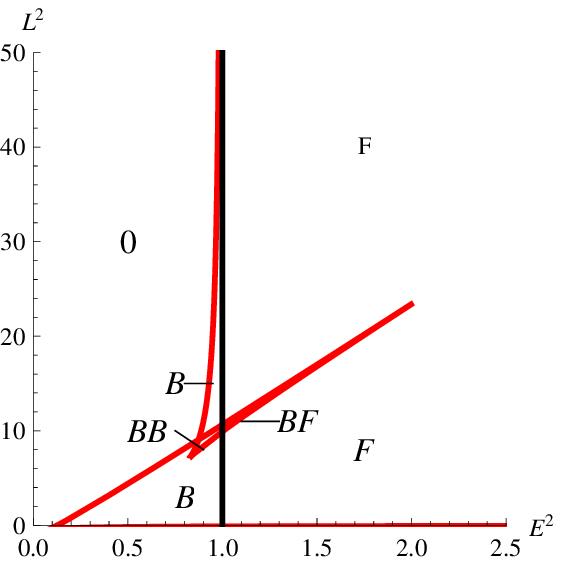}}\\
\subfloat[(e) $\bar Q = 1.09$]{\includegraphics[width=0.42\textwidth]{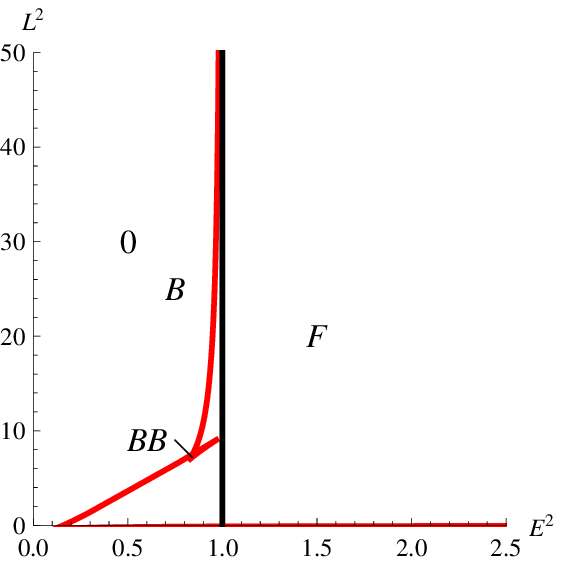}}\quad
\subfloat[(f) $\bar Q = 1.5 > \bar Q_{\rm cr}$]{\includegraphics[width=0.42\textwidth]{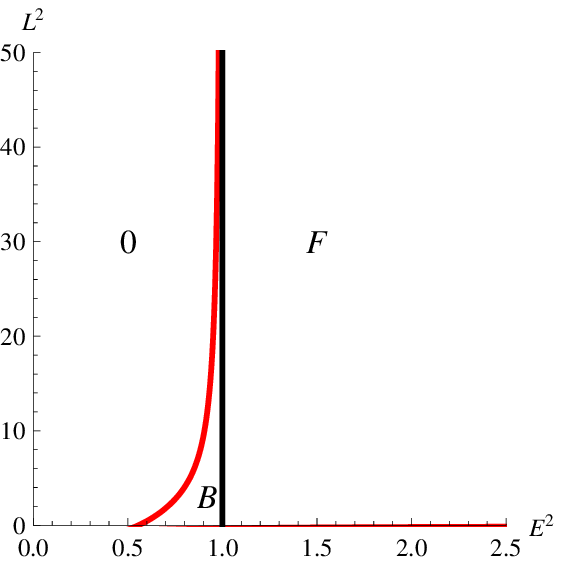}}
\caption{Orbit types for massive neutral test particles in Reissner-Nordstr\"om space-times (in Schwarzschild space-times we also have terminating orbits $T$ falling into the singularity). The black vertical lines mark $E=1$, and the other red solid lines mark circular orbits. They divide different types of orbit configurations.}
\label{Fig:RNMassive}
\end{figure}

For the sake of a better comparison we present here the orbits in Reisser-Nordstr\"om space--times. The metric in these space-time is given by \eqref{generalmetric} with 
\begin{equation}
g_{tt} = \frac{1}{g_{rr}} = 1 - \frac{2 M}{r} + \frac{Q^2}{r^2} \, .
\end{equation}
From \eqref{drdphi} we obtain the orbital equation
\begin{equation}
\left(\frac{dr}{d\varphi}\right)^2 = \frac{r^4}{L^2} \left(E^2 - \left(1 - \frac{2 M}{r} + \frac{Q^2}{r^2}\right) \left(\epsilon + \frac{L^2}{r^2}\right)\right) \, ,
\end{equation}
where $E$ and $L$ are the conserved energy and angular momentum. the effective potential reads
\begin{equation}
V_{\rm eff} = - \epsilon \frac{M}{r} + \frac{L^2 + Q^2}{r^2} - \frac{M L^2}{r^3} + \frac{Q^2 L^2}{r^4} \, .
\end{equation}
A substitution $u = 2 m/r$ then gives for the orbital equation 
\begin{equation}
\left(\frac{du}{d\varphi}\right)^2 = \frac{4}{\bar L^2} \left(E^2 - \epsilon\right) + \epsilon \frac{4}{\bar L^2} u - \left(1 + \epsilon \frac{\bar Q^2}{\bar L^2}\right) u^2 + u^3 - \frac{\bar Q^2}{4} u^4 \, ,
\end{equation}
where we used the normalized quantities defined in Sec.~3.1. This equation has the form 
\begin{equation}
\left(\frac{du}{d\varphi}\right)^2 = P_4(u) \, ,
\end{equation}
where $P_4$ is a polynomial of order 4. The structure of orbits is given by the zeros of the polynomial which depend on $E^2$, $\bar L^2$, and $\bar Q^2$. The number of zeros and the corresponding types of orbits are given by the parameter plots of Fig.~\ref{Fig:RNMassive}. This has to be compared with Fig.~\ref{Fig:Massive}.

Fig.~\ref{Fig:RNMassive}(a) describes the manifold of orbits in Schwarschild space-times \cite{Hagihara31}. Fig.~\ref{Fig:RNMassive}(b) shows all orbits in a proper Reissner-Nordstr\"om black hole space-time with small charges as to allow the existence of two horizons. Here we have (i) two bound orbits where one crosses both horizons, (ii) one bound orbit crossing both horizons together with a flyby orbit, and (iii) a bound orbit crossing again both horizons. Bound orbits not crossing the horizons are possible only in the small region left to the vertical line. 

The orbits in an extremal Reissner-Nordstr\"om space-time are shown in Fig.~\ref{Fig:RNMassive}(c). In this case we have (i) two bound orbits, (ii) one bound orbit, or (iii) one bound and one flyby orbit.  

Figs.~\ref{Fig:RNMassive}(d) and (e) show orbits in a regular Reissner-Nordstr\"om space-time where the charge is small enough in order to allow two bound orbits which appear for parameters in the small region left from the vertical line. 

\begin{figure}[t!]
\begin{center}
\includegraphics[width=7cm]{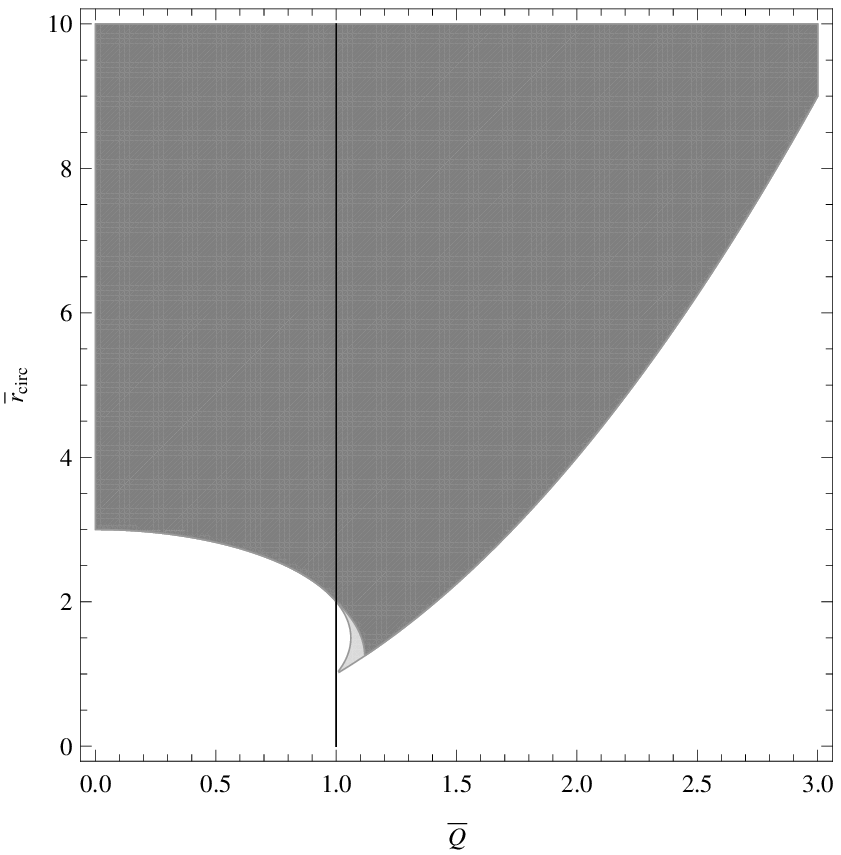}
\end{center}
\caption{Radii of circular orbits for neutral test particles in Reissner-Nordstr\"om space-times as a function of the charge $\bar Q$. The dark gray region corresponds to stable orbits and the light gray region to unstable orbits. \label{RNCircularOrbits}}
\end{figure}

In Reissner-Nordstr\"om space-time circular orbits are given by the conditions $P_4(u) = 0$ and $\frac{dP_4}{du}(u) = 0$ what can be solved for $E^2$ and $\bar L^2$
\begin{equation}
\begin{aligned}
E^2 & = \frac18 \frac{\left(4 - 4 u + \bar Q^2 u^2\right)^2}{2 - 3 u + \bar Q^2 u^2} \\
\bar L^2 & = 2 \frac{2 - \bar Q^2 u}{u \left(2 - 3 u + \bar Q^2 u^2\right)} \, .
\end{aligned}
\end{equation}
Since both $E^2$ and $\bar L^2$ have to be positive, the necessary conditions or the existence of circular orbits are $2 - 3 u + \bar Q^2 u^2 > 0$ and $2 - \bar Q^2 u > 0$ for $u > 0$. A circular orbit is stable if the second derivative of $P_4$ is negative, $\frac{d^2 P_4}{du}(u) < 0$. These condition together give Fig.~\ref{RNCircularOrbits} where the light gray region shows unstable circular orbits, and the dark gray region stable circular orbits. This Figure has to be compared with Fig.~\ref{Fig:ISCO}.  

~

~

\end{document}